%% file: root.tex
\documentclass[journal,twoside,web]{ieeecolor}
\usepackage{tmi}
\usepackage{cite}
\usepackage{amsmath,amssymb,amsfonts}
\usepackage{graphicx}
\usepackage{textcomp}
\usepackage{xspace}
\usepackage{bm}
\input{macros}
\usepackage{multirow}
\usepackage{float}
\usepackage{hyperref}
\usepackage{algorithm}
\usepackage{algpseudocode}
\usepackage{slashbox}

\def\BibTeX{{\rm B\kern-.05em{\sc i\kern-.025em b}\kern-.08em
    T\kern-.1667em\lower.7ex\hbox{E}\kern-.125emX}}
\begin{document}

\title{B-spline Parameterized Joint Optimization of Reconstruction
and K-space Trajectories (BJORK)
for Accelerated 2D MRI}
\author{Guanhua Wang, Tianrui Luo, Jon‐Fredrik Nielsen,
Douglas C. Noll, \IEEEmembership{Senior Member, IEEE}, and Jeffrey A. Fessler \IEEEmembership{Fellow, IEEE}
\thanks{
This work is supported in part by
NIH Grants R01 EB023618 and U01 EB026977,
and NSF Grant IIS 1838179.}
\thanks{G. Wang, T. Luo, J.-F. Nielsen, and D. C. Noll are with the
Department of Biomedical Engineering, University of Michigan, Ann Arbor, MI 48109 USA (e-mails:\{guanhuaw, tianrluo, jfnielse, dnoll\}@umich.edu).}
\thanks{J. A. Fessler is with the Department of EECS, University of Michigan, Ann Arbor, MI 48109 USA (e-mail:fessler@umich.edu).}}
\maketitle

\begin{abstract}
Optimizing k-space sampling trajectories
is a promising yet challenging topic
for fast magnetic resonance imaging (MRI).
This work proposes to optimize a reconstruction method
and sampling trajectories jointly
concerning image reconstruction quality
in a supervised learning manner.
We parameterize trajectories with quadratic B-spline kernels 
to reduce the number of parameters
and apply multi-scale optimization,
which may help to avoid sub-optimal local minima.
The algorithm includes an efficient non-Cartesian
unrolled neural network-based reconstruction
and an accurate approximation for backpropagation through
the non-uniform fast Fourier transform (NUFFT) operator
to accurately reconstruct and back-propagate
multi-coil non-Cartesian data.
Penalties on slew rate and gradient amplitude
enforce hardware constraints.
Sampling and reconstruction are trained jointly
using large public datasets. 
To correct for possible eddy-current effects
introduced by the curved trajectory,
we use a pencil-beam trajectory mapping technique.
In both simulations and in-vivo experiments,
the learned trajectory demonstrates
significantly improved image quality
compared to previous model-based and
learning-based trajectory optimization methods
for 10$\times$ acceleration factors.
Though trained with neural network-based reconstruction,
the proposed trajectory also leads to improved image quality
with compressed sensing-based reconstruction. 
\end{abstract}


\section{Introduction}
\label{sec:intro}
\input{s,intro}

\section{Methods}
\label{sec:methods}
\input{s,method}

\section{Experiments}
\label{sec:experiments}
\input{s,exp}

\section{Results}
\label{sec:results}
\input{s,res}

\section{Discussion}
\label{sec:discussion}
\input{s,dis}

\pagebreak
\appendix

\input{s,supp}
% \clearpage
\section*{Acknowledgment}
The authors thank Dr. Melissa Haskell
and Naveen Murthy
from University of Michigan
for helpful advice and fruitful discussions.
We also thank Dr. Matthew Muckley for 
the PyTorch-based NUFFT toolboxes\footnote{\url{https://github.com/mmuckley/torchkbnufft}} \cite{muckley:20:tah}.

% \vfill
% \pagebreak

\bibliographystyle{IEEEtran}
\bibliography{refs}

\clearpage

\end{document}

%% file: macros.tex
\long\def\comment#1{}

\newcommand{\fref}[1] {Fig.~\ref{#1}\xspace} % the non-breaking space "~" is very important!  use macros!
\newcommand{\tref}[1] {Table~\ref{#1}\xspace}

\newcommand{\xmath}[1] {\ensuremath{#1}\xspace}
\newcommand{\blmath}[1] {\xmath{\bm{#1}}}

\newcommand{\norminfs}[1] {\xmath{\| #1 \|_{\infty}}}

\newcommand{\resp}[1]{}

\newcommand{\omdd}[1] {\xmath{\bm{\omega}^{[#1]}}}
\newcommand{\omd} {\omdd{d}}

\newcommand{\cc} {\xmath{\bm{c}}}
\newcommand{\cd} {\xmath{\bm{c}^{[d]}}}

\newcommand{\argmin} {\operatornamewithlimits{arg\,min}} % argmin

\newcommand{\reals} {\xmath{\mathbb{R}}}
\newcommand{\complex} {\xmath{\mathbb{C}}}

\newcommand{\A} {\blmath{A}}
\newcommand{\B} {\blmath{B}}
\newcommand{\D} {\blmath{D}}
\newcommand{\I} {\blmath{I}}
\newcommand{\R} {\blmath{R}}

\newcommand{\x} {\blmath{x}}
\newcommand{\xh} {\xmath{\hat{\x}}}

\newcommand{\y} {\blmath{y}}

\newcommand{\vveps} {\blmath{\varepsilon}}
\newcommand{\om} {\blmath{\omega}}
\newcommand{\thta} {\blmath{\theta}}
\newcommand{\z} {\blmath{z}}

\newcommand{\gmax} {\xmath{g_{\max}}}
\newcommand{\smax} {\xmath{s_{\max}}}

\newcommand{\Ns} {\xmath{N_{\mathrm{s}}}} % # samples
\newcommand{\Nd} {\xmath{N_{\mathrm{d}}}} % # dimensions
\newcommand{\Nv} {\xmath{N_{\mathrm{v}}}} % # voxels
\newcommand{\Nc} {\xmath{N_{\mathrm{c}}}} % # coils
\newcommand{\Nlevel} {\xmath{N_{\mathrm{level}}}}
\newcommand{\Nepoch} {\xmath{N_{\mathrm{epoch}}}}
\newcommand{\Niter} {\xmath{N_{\mathrm{iter}}}}

\newcommand{\cD} {\xmath{\mathcal{D}}}

\long\def\red#1{\bgroup\color{red} #1 \egroup}
\long\def\blue#1{ #1} % this version allows par breaks!

%% file: s,intro.tex
Magnetic Resonance Imaging (MRI) systems
acquire raw data in the frequency domain (k-space).
Most scanning protocols sample data points sequentially
according to a pre-determined sampling pattern.
The most common sampling patterns are
variants of Cartesian rasters
and non-Cartesian trajectories such as radial spokes \cite{lauterbur1973image}
and spiral interleaves \cite{ahn1986spiral}.
The local smoothness of these patterns
facilitates ensuring that they obey hardware limits,
namely the maximum gradient and slew rate
that constrain the speed and acceleration
when traversing k-space.
These patterns also make it easy to ensure
sufficient sampling densities.
In recent years, hardware improvements,
especially with the RF and gradient systems,
enable more complex gradient waveform designs and sampling patterns.
For a given readout time,
optimized designs can cover a broader 
and potentially more useful region in k-space,
reducing the overall scanning time 
and/or improving image quality,
particularly when combined with multiple receive coils.

For fast imaging, many works focus on
acceleration in the phase-encoding (PE) direction
with fully sampled frequency-encoding (FE) lines 
\cite{larkman:2007:ParallelMagneticResonance, wang:2010:VariableDensityCompressed,
knoll:2011:AdaptedRandomSampling, seeger:2010:OptimizationKspaceTrajectories,
chauffert:2013:VariableDensityCompressed}.
Usually, there is enough time for the $\Delta k$
shifts in the PE direction,
so gradient and slew rate constraints
are readily satisfied.
More general non-Cartesian 
trajectory designs in 2D and 3D
can further exploit the flexibility in the FE direction.
However, in addition to hardware physical constraints, 
MRI systems are affected by imperfections
such as the eddy currents 
that cause the actual trajectory to deviate from the nominal one
and introduce undesired phase fluctuations
in the acquired data \cite{robison:2019:CorrectionB0Eddy}.
Some studies optimize properties of existing trajectories
such as the density of spiral trajectories \cite{lee:2003:Fast3DImaging}
or the rotation angle of radial trajectories \cite{winkelmann:2007:OptimalRadialProfile}.
More complex waveforms,
e.g., wave-like patterns \cite{bilgic:2015:WaveCAIPIHighlyAccelerated},
can provide more uniform coverage of k-space 
and mitigate aliasing artifacts.
To accommodate the incoherence requirements
of compressed sensing based methods,
\cite{bilgin:08:random, lustig:2008:FastMethodDesigning}
introduce slight perturbations to existing trajectories,
like radial or spiral trajectories.
Some works also explore genetic algorithms
to solve this non-convex constrained problem
\cite{sabat:2003:ThreeDimensionalKspace}. 

The recent SPARKLING method \cite{sparklingmrm, weiss2021optimizing, lazarus20203d}
considers two criteria for trajectory design: 
(1) the trajectory should match a pre-determined sampling density
according to a certain measure, 
and (2) the sampling points should be
locally uniform to avoid clusters or gaps. 
The density and uniformity criteria are transformed
into ``attraction'' and ``repulsion'' forces among the sampling points. 
The work uses fast multipole methods (FMM) \cite{fong:2009:BlackboxFastMultipole} to
efficiently calculate the interactions between points. 
Projection-based optimization
handles the gradient and slew rate constraints
\cite{chauffert:2016:ProjectionAlgorithmGradient}.
In-vivo and simulation experiments demonstrate
that this approach reduces
aliasing artifacts for 2D and 3D T2*-weighted imaging. 
However, in SPARKLING,
the density is determined heuristically;
determining the optimal sampling density
for different protocols remains an open problem. 
The work also does not consider some k-space signal characteristics
such as conjugate symmetry. 
Furthermore, the point spread function (PSF)
of the calculated trajectory for high under-sampling rates
may be suboptimal for
reconstruction algorithms like those based on convolution neural networks,
because the reconstruction algorithm
is not part of the SPARKLING design process.

With rapid advances in deep learning 
and auto-differentiation software, 
learning-based signal sampling strategies are being 
investigated in multiple fields such as 
optics and ultrasound
\cite{elmalem:2018:LearnedPhaseCoded, huijben:2020:LearningSubSamplingSignal}. 
In MRI, most learning-based works
have focused on sampling patterns of phase encoding locations. 
Some studies formulate the on-grid sampling pattern 
as i.i.d samples from multivariate Bernoulli distribution 
\cite{bahadir:2020:DeepLearningBasedOptimizationUnderSampling, huijben:2020:LearningSamplingModelBased}. 
Since random sampling operations are not differentiable, 
different surrogate gradients, such as Gumbel-Softmax,
are developed in these works. 
Rather than gradient descent,
\cite{sanchez:2020:ScalableLearningBasedSampling} uses a greedy search method. 
\cite{zibetti2020fast} further reduces the complexity of greedy search by
Pareto optimization, an evolutionary algorithm 
for sparse regression \cite{qian:2015:SubsetSelectionPareto}.
Some works have used reinforcement learning. 
For example, \cite{jin:2019:SelfSupervisedDeepActive} and \cite{rl:david}
adopted a double network setting: 
one for reconstruction and the other generating a sampling pattern, 
where the first work used Monte-Carlo Tree Search (MCTS) 
and the second used Q-learning to optimize
the 1-D sub-sampling.
Instead of using an end-to-end CNN
as the reconstruction algorithm in other works,
\cite{sherry:20:lts} constructs a differentiable
compressed sensing reconstruction framework. 
\cite{aggarwal:2020:JointOptimizationSampling} used an unrolled neural network
as the reconstruction algorithm.

To our knowledge, PILOT \cite{pilot} is the first work
to optimize a 2D non-Cartesian trajectory
and an image reconstruction method simultaneously.
The training loss is the reconstruction error
since the ultimate goal of trajectory optimization is high image quality.
The trained parameters
were the locations of sampling points
and the weights of the reconstruction neural network.
Large datasets and stochastic gradient descent
were used to optimize the parameters.
To meet the hardware limits,
a penalty was applied on the gradient and slew rate.
Since the reconstruction involves non-Cartesian data,
PILOT uses a
(bilinear, hence differentiable almost everywhere)
gridding reconstruction algorithm
to map the k-space data into the image domain,
followed by a U-Net \cite{unet} to refine the gridded image data.
Simulation experiments report encouraging results
compared to ordinary trajectories.
Nevertheless, the algorithm often gets stuck
in sub-optimal local minima
where the initial trajectory is only slightly perturbed
yet the slew rate rapidly oscillates.
To reduce the effect of initialization,
\cite{pilot}
uses a randomized initialization algorithm based on the
traveling salesman problem (TSP).
However, this initialization approach works
only with single-shot long TE sequences,
limiting its utility in many clinical applications.
The implementation in
\cite{pilot}
relies on auto-differentiation
to calculate the Jacobian of the non-uniform Fourier transform;
here we adopt a new NUFFT Jacobian approximation
that is faster
and more accurately approximates
the non-Cartesian discrete Fourier transform (DFT)
\cite{wang:21:eao}.

To overcome the limitations of previous methods
and further expand their possible applications,
this paper proposes an improved
supervised learning workflow
called \textbf{B}-spline parameterized \textbf{J}oint
\textbf{O}ptimization of \textbf{R}econstruction
and \textbf{K}-space trajectory (\textbf{BJORK}).
Our main contributions include the following.
(1) We parameterize the trajectories
with quadratic B-spline kernels.
The B-spline reparameterization reduces the number of parameters
and facilitates multilevel optimization,
enabling non-local improvements
to the initial trajectory.
Moreover, the local smoothness of B-spline kernels
avoids rapid waveform oscillations.
(2) We adopt an unrolled neural network reconstruction method
for non-Cartesian sampling patterns \cite{modl}.
Compared to the image-domain U-Net implemented in previous works,
the proposed approach combines the strength of learning-based
and model-based reconstruction,
improving the effect of both
reconstruction and trajectory learning.
(3) We adopt accurate and efficient NUFFT-based approximations
of the Jacobian matrices of the DFT operations used
in the reconstruction algorithm.
(See \cite{wang:21:eao} for detailed derivations and validation.)
(4) In addition to a simulation experiment,
we also conducted phantom and in-vivo experiments
with protocols that differ from the training dataset
to evaluate the generalizability and applicability of the model.
(5) We used a k-space mapping technique
to correct potential eddy current-related artifacts.
(6) Compared with SPARKLING, the proposed learning-based approach
does not need to assume signal characteristics
such as spectrum energy density.
Instead,
BJORK learns the required sampling trajectories
from a large data set
in a supervised manner.

The remaining materials are organized as follows.
Section~II details the proposed method.
Section~III describes experiment settings and control methods.
Sections~IV and~V present and discuss the results.

\begin{figure*}[htbp!]
    \centering
    \includegraphics[width=0.95\textwidth]{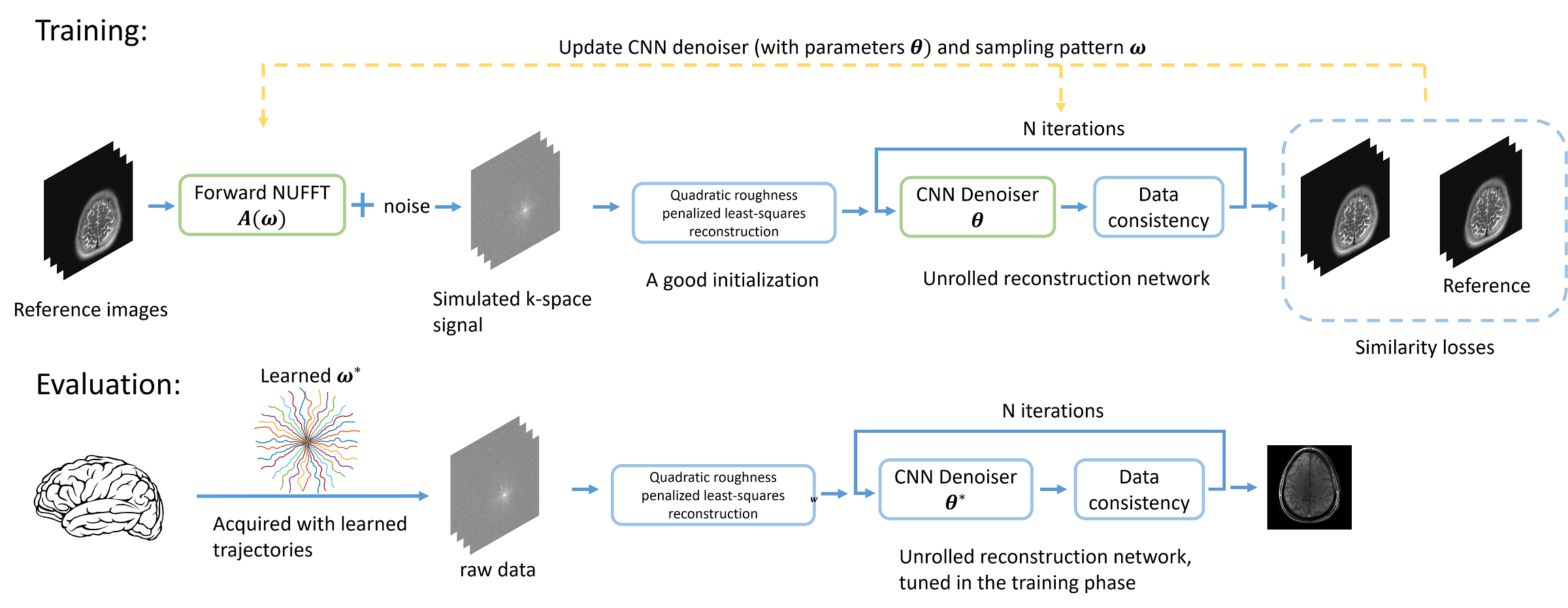}
    \caption{Diagram of the proposed approach.
    To optimize the sampling trajectory
    and the reconstruction algorithm jointly
    using a stochastic gradient descent (SGD)-type method, 
    we construct a differentiable forward MRI system matrix
    $\A(\om)$ that simulates k-space data
    w.r.t. trajectory \om from ground truth images,
    and an unrolled neural network
    for reconstruction.
    The reconstruction errors compared with the ground truth
    are used as the training loss to update learnable parameters
    (the trajectory \om and the network's parameters \thta).}
    \label{fig:workflow}
\end{figure*}

%% file: s,method.tex
% s,method

This section describes the proposed approach
for supervised learning
of the sampling trajectory
and image reconstruction method.

\subsection{Problem formulation}
\label{subsec:problem}

\fref{fig:workflow} shows the overall workflow of the proposed approach.
The goal is to optimize
$\om \in \reals^{\Ns \times \Nd}$,
a trainable (possibly multi-shot) sampling pattern,
and
$\thta \in \reals^M$,
the $M$ parameters of the image reconstruction method,
where
\Ns denotes the total number of k-space samples,
and \Nd denotes the image dimensionality.
(The results are for $\Nd = 2$,
i.e., 2D images,
but the method is general.)

The training loss
for jointly optimizing the trajectory parameters \om
and reconstruction parameters \thta
is as follows:
\begin{align}
\label{jointopt}
%\MoveEqLeft
&\argmin_{\om \in \reals^{\Ns \times \Nd},\,
\thta \in \reals^M} \mathbb{E}_{\x \in \mathcal{X}}[
\ell(f_{\thta}(\om; \A(\om)\x+\vveps), \x) ]
\\
\text{s.t. }\ 
& \norminfs{ \D_1 \omd } \leq \gamma \Delta t \gmax
, \nonumber \\
& \norminfs{ \D_2 \omd } \leq \gamma \Delta t^2 \smax
, \  d = 1,\ldots,\Nd, \nonumber
\end{align}
%}
%\blue{
where each
$\x \in \complex^{\Nv}$
is a
%(batch of)
fully sampled reference image
having \Nv voxels
drawn from the training data set $\mathcal{X}$
and \vveps is simulated additive complex Gaussian noise.
(In practice the expectation is taken
over mini-batches of training images.)

The system matrix
$\A = \A(\om) \in \complex^{\Ns\Nc \times \Nv}$
represents the MR imaging physics
(encoding),
where \Nc denotes the number of receiver coils. 
For multi-coil non-Cartesian acquisition, 
it is a non-Cartesian SENSE operator \cite{pruessmann:2001:AdvancesSensitivityEncoding} 
that applies a pointwise multiplication of the sensitivity maps 
followed by a NUFFT operator
(currently we do not consider field inhomogeneity but it would be straightforward to extend because the Jacobian approximation can cover such cases \cite{wang:21:eao}).
The function
$f_{\thta}(\om;\cdot)$ denotes an image
reconstruction algorithm
with parameters $\thta$
that is applied to simulated under-sampled data
$\A(\om)\x + \vveps$.
As detailed in subsection \ref{subsec:recon},
we use an unrolled neural network.
The reconstruction loss
%$\|\cdot\|$
$\ell(\cdot, \cdot)$
quantifies the similarity between a reconstructed image and the ground truth,
and can be a combination of different terms.
Here we chose the loss $\ell$
to be a combined $\ell_1$ and square of $\ell_2$ norm.
The matrices $\D_1$ and $\D_{2}$
denote the first-order and second-order finite difference operators.
$\Delta t$ is the raster time
and $\gamma$ denotes the gyromagnetic ratio.
\resp{R4.12}
\resp{R4.13}
For multi-shot imaging,
the difference operator applies to each shot individually.
The optimization is constrained in gradient field strength (\gmax),
and slew rate (\smax).
To use the stochastic gradient descent (SGD) method,
we convert the box constraint into
a penalty function $\blmath{\phi}$,
where
\[
\blmath{\phi}_\lambda(|\x|)
= \bm{1}^T \max.(|\x|-\lambda, 0),
\]
where $\max.(\cdot)$ operates point-wisely.
Our final joint optimization problem has the following form:
\begin{align}
\label{objective}
\argmin_{\om \in \mathbb{C}^{\Ns \times \Nd},\,
\blmath{\theta} \in \reals^{M}}\mathbb{E}_{\x \in \mathcal{X}}[ &
\, \ell( f_{\blmath{\theta}, \om}(\om; \A(\om)\x + \vveps), \x ) ]
\\
& + \mu_1 \blmath{\phi}_{\gamma \Delta t \gmax}(|\D_1\om|)\nonumber \\ 
& + \mu_2 \blmath{\phi}_{\gamma\Delta t^{2} \smax}(|\D_2\om|)\nonumber.
\end{align}
We update $\thta$ and $\om$ simultaneously
for each mini-batch of training data.

\subsection{Parameterization and multi-level optimization}
\label{subsec:param}
We parameterize the sampling pattern with 2nd-order quadratic B-spline kernels: 
\begin{equation}
\omd = \B \cd
,\ d=1,\ldots,\Nd,
\label{parameterization}
\end{equation}
where $\B \in \reals^{\Ns \times L}$
denotes the interpolation matrix,
and $\cd$ denotes
the $d$th column of the coefficient matrix
$\cc \in \reals^{L \times \Nd}$.
$L$ denotes the length of $\cd$,
or the number of interpolation kernels in each dimension.
The decimation rate in \fref{evo} is defined as
$\text{Decim.} = \Ns/L$.
Compared to other parameterization kernels,
B-spline kernels reduce
the number of individual inequality constraints
(on maximum gradient strength 
and slew rate)
from $4 \Nd \Ns$ to $4 \Nd L$
where typically $L \ll \Ns$.
See \cite{hao:2016:JointDesignExcitationa} for more details.

Early versions of previous work \cite{pilot}
and our preliminary experiments
found optimized trajectories
that were often local minima near the initialization,
only slightly perturbing the initial trajectory%
\footnote{
The latest versions of PILOT on arXiv
\cite[versions 4-5]{pilot}
also use trajectory parameterization,
focusing on long readout time cases.
}.
We use a multilevel training strategy
to improve the optimization process
\cite{Nielsen:16:Improved, boyer2016generation}.

We initialized
the decimation rate $\Ns/L$
with a large value (like 64).
Thus, many neighboring sample points
are controlled by the same coefficient,
which introduces more `non-local' improvements.
After both \cc and \thta converge,
we reduce the decimation rate,
typically by a factor of $2$,
and begin a new round of training
initialized with \om and \thta of the previous round. 
\fref{evo} depicts the evolution along with decimation rates.

\subsection{Reconstruction}
\label{subsec:recon}

In the joint learning model,
we adopted a model-based unrolled neural network (UNN) approach 
to image reconstruction
\cite{modl, yang:2016:DeepADMMNetCompressivea, hammernik:2018:LearningVariationalNetwork, schlemper:2019:SigmaNetEnsembled}.
Compared to the previous joint learning model (PILOT)
that used a single image domain %end-to-end
network \cite{pilot},
an unrolled network can lead to a more accurate reconstruction
\cite{modl},
at the price of longer reconstruction time.

A typical cost function
for regularized MR image reconstruction
has the form:
\begin{equation}
\label{recon}
\xh = \argmin_{\x} \|\A\x-\y\|_{2}^2 + \mathcal{R}(\x).
\end{equation}
The first term is usually called the data-consistency term
that ensures the reconstructed image 
is consistent with the acquired k-space data \y.
(In the training phase,
$\A(\om)\x+\vveps$ is the simulated \y.)
The regularization term
$\mathcal{R}(\cdot)$
is designed to control aliasing and noise
when the data is under-sampled.
By introducing an auxiliary variable $\z$,
one often replaces \eqref{recon}
with the following alternative:
\begin{equation}
\label{recon_aux}
\xh = \argmin_{\x} \min_{\z} \|\A\x-\y\|_2^2
+ \mathcal{R}(\z) + \mu\|\x-\z\|_{2}^2,
\end{equation}
where $\mu > 0$ is a penalty parameter. 
Using an alternating minimization approach,
the optimization updates become:
\begin{equation}
\label{x1}
\x_{i+1} = \argmin_{\x} {\|\A\x-\y\|_{2}^2 + \mu\|\x-\z_{i}\|_{2}^2},
\end{equation}
\begin{equation}
\label{z1}
\z_{i+1} = \argmin_{\z}{\mathcal{R}(\z) + \mu\|\x_{i+1}-\z\|_{2}^2}.
\end{equation}
The analytical solution for the \x update is
\[
\x_{i+1} = (\A'\A+ \mu \I)^{-1} (\A'\y + \mu \z_i),
\]
which involves
a matrix inverse
that would be
computationally prohibitive
to compute directly.
Following \cite{modl},
we use a few iterations of the conjugate gradient (CG) method 
for the \x update.
The implementation uses a Toeplitz embedding technique
to accelerate the computation of $\A'\A$ \cite{fessler:05:tbi, muckley:20:tah}.

For a mathematically defined regularizer,
the \z update would be a proximal operator.
Here we follow previous work
\cite{gregor:2010:LearningFastApproximations,modl}
and use a CNN-based denoiser
$
\z_{i+1} = \cD_{\thta}(\x_{i+1})
$.
To minimize memory usage and avoid over-fitting,
we used the same \thta across iterations,
though iteration-specific networks may
improve the reconstruction result \cite{schlemper:2019:SigmaNetEnsembled}.

For the CNN-based denoiser,
we used the Deep Iterative Down-Up CNN (DIDN) \cite{DIDN,schlemper:2019:SigmaNetEnsembled}.
As a state-of-art model for image denoising,
the DIDN model requires less memory than popular models like U-net \cite{unet}
while providing improved reconstruction results.
In our experiments, it also led to faster
training convergence
than previous denoising networks.

Since neural networks are sensitive to the scale of the input,
a good and consistent initial estimate of $\x$ is important.
We used the following quadratic roughness penalty approach
to compute an initial image estimate:
\begin{align}
\label{init}
\x_{0} &= \argmin_{\x}{\|\A\x-\y\|_{2}^2
+ \lambda\|\R \x\|_{2}^2}\\
&= (\A'\A+ \lambda \R'\R)^{-1}\A'\y, \nonumber
\end{align}
where \R denotes the \Nd-dimensional first-order finite difference (roughness) operator.
We also used the CG method to (approximately) solve
this quadratic minimization problem.

\subsection{Correction of eddy-current effect}
\label{subsec:eddy}

Rapidly changing gradient waveforms
may suffer from eddy-current effects,
even with shielded coils.
This hardware imperfection requires additional measurements and corrections
so that the actual sampling trajectory
is used for reconstructing real MRI data.
Some previous works used a field probe
and corresponding gradient impulse-response (GIRF) model \cite{vannesjo:2013:GradientSystemCharacterization}.
In this work, we adopted the `k-space mapping' method \cite{duyn:1998:SimpleCorrectionMethod,robison:2019:CorrectionB0Eddy}
that does not require additional hardware.
Rather than mapping the $k_x$ and $k_y$ components 
separately as in previous papers,
we excited a pencil-beam region using one $90^{\circ}$ flip
and a subsequent $180^{\circ}$ spin-echo pulse
\cite{nielsen:2018:TOPPEFrameworkRapid}.
We averaged multiple repetitions to estimate
the actual acquisition trajectory.
We also subtracted
a zero-order eddy current phase term
from the acquired data \cite{robison:2019:CorrectionB0Eddy}.

The following pseudo-code
summarizes the BJORK training process.
\input{a,bjork}

%% file: a,bjork.tex
% \begin{algorithm}
% \caption{The algorithm for BJORK}
% \label{alg:bjork}
% \begin{algorithmic}
% \Require denoiser $\cD_{\thta}$, trajectory $\om$
% % \Ensure
% \
% \State $\thta \gets \thta_0$
% \State $\om \gets \om_0$
% \State Pre-train $\cD_{\thta}$ with fixed $\om$.
% \For 
% \If{$N$ is even}
%     \State $X \gets X \times X$
%     \State $N \gets \frac{N}{2}$  \Comment{This is a comment}
% \ElsIf{$N$ is odd}
%     \State $y \gets y \times X$
%     \State $N \gets N - 1$
% \EndIf
% \EndFor
% \end{algorithmic}
% \end{algorithm}
% \algnewcommand\INPUT{\State \textbf{Inputs: }}
% \algnewcommand\DATA{\State \textbf{Data: }}
% \algnewcommand\algindent{\hspace{\algorithmicindent}}
\algnewcommand\INPUT{\item[\textbf{Input:}]}%
\algnewcommand\Output{\item[\textbf{Output:}]}%
\algnewcommand\DATA{\item[\textbf{Data:}]}%

\begin{algorithm}[H]
  \caption{Training algorithm for BJORK}
  \label{alg:bjork}
  \begin{algorithmic}[1]
    \Require Training set $\mathcal{X}$;
    denoiser $\cD_{\thta}$
    for initial CNN weights $\thta_0$;
    initial trajectory $\om_0$;
    levels of optimization \Nlevel;
    number of epoch \Nepoch;
    step size of denoiser update $\eta_{\cD}$;
    step size of trajectory update $\eta_{\om}$;
    penalty parameter for gradient/slew rate constraint $\mu_1$ and $\mu_2$.
    \Ensure $\om = \B \cc$
    \State $\thta \gets \thta_0$
    \State $\om \gets \om_0$
    \State Pre-train $\cD_{\thta}$ with fixed $\om_0$.
    \For{$l$ = 1 to \Nlevel} % we use $i$ for MODL iteration!!
    \State Initialize new coefficient matrix $\B_l$.
    \State Initialize new coefficient $\cc_l^0$ with $\om_{l-1}\approx \B_l \cc_l^0$.
%    \Comment{Solve by inverse problem}
    \For{$j$ = 1 to \Nepoch}
        \For{training batches $\x^K$ in $\mathcal{X}$}
        \State \textbf{Simulate the k-space w.r.t. $\om_l$:}
        \State $\y^K = \A(\om_l^K) \x^K +\vveps$
        \State \textbf{Reconstruction with UNN:}
        \State Reconstruct initial images using \eqref{init} with CG
        \For{$i$ = 1 to \Niter}
            \State $\x_{i+1}$: UNN reconstruction update of $\z_{i}$
            \State \hspace*{1em} using \eqref{x1}
            \State Apply CNN: $\z_{i+1} = \cD_{\thta}(\x_{i+1})$
        \EndFor
        %\State $\xh^K = \argmin_{\x} \|\A\x-\y^K\|_{2}^2 + \mathcal{R}(\x)$ % no! no R!
        \State \textbf{Calculate loss function:}
        \State $L = \ell(\xh^K, \x^K)
        + \mu_1 \blmath{\phi}_{\gamma \Delta t \gmax}(|\D_1\om_i^K|)$
        \State $\quad \mbox{}
        + \mu_2 \blmath{\phi}_{\gamma\Delta t^{2} \smax}(|\D_2\om_i^K|)$
        \State \textbf{Update denoiser and trajectory:}
        \State $\thta^K = \thta^{K-1} - \eta_{\cD} \nabla_{\thta^{K-1}}L $
        \State $\om_l^K = \om_l^{K-1} - \eta_{\om} \nabla_{\om_l^{K-1}}L $
        \EndFor   
    \EndFor
\EndFor
\end{algorithmic}
\end{algorithm}
\vspace{-1.5\baselineskip}

%% file: s,exp.tex
% s,exp

\input{t,proto}

\subsection{Comparison with prior art}
\label{subsec:comp}

We compared the proposed BJORK approach
with the SPARKLING method for trajectory design
in all experiments,
and have set the readout length and physical constraints to be the same
for both methods.
% SPARKLING used the default multi-level optimization strategy 
% and parameter settings, 
% as detailed in \cite{lazarus2018compressed}.

Both BJORK and PILOT \cite{pilot}
are methods for joint sampling design and reconstruction optimization.
We compared three key differences between 
the two methods individually.
(1) The NUFFT Jacobian matrices,
as discussed in \cite{wang:21:eao} and the Appendix.
(2)
The reconstruction method involved.
Our BJORK approach uses an unrolled neural network,
while PILOT uses a single 
%end-to-end
reconstruction neural network
in the image domain (U-Net).
We also presented 
the effect of
trajectory parameterization
(BJORK uses quadratic B-splines following
\cite{hao:2016:JointDesignExcitationa},
whereas versions 1-3 of PILOT
used no parameterization
and more recent versions of PILOT
use cubic splines
\cite{pilot}).

\subsection{Image quality evaluation}
\label{subsec:IQ}

To evaluate the reconstruction quality
provided by different trajectories,
we used two types of reconstruction methods
in the test phase:
unrolled neural network (UNN)
(with learned \thta)
and a compressed sensing approach
(sparsity regularization for an discrete wavelet transform).
For SPARKLING-optimzed trajectories and standard undersampled trajectories (radial/spiral), 
we used the same unrolled neural networks as in BJORK for reconstruction.
Only the network parameters \thta were trained,
with the trajectory \om fixed.

We also used compressed sensing-based reconstruction
to test the generalizability of BJORK-optimized trajectories.
The penalty function is the $\ell_1$ norm of a discrete wavelet transform
with a Daubechies 4 wavelet.
The ratio between the penalty term and the data-fidelity term is
$10^{-7}$. %1e-7.
We used the SigPy package\footnote{\url{https://github.com/mikgroup/sigpy}}
and its default primal-dual hybrid gradient (PDHG) algorithm (with 50 iterations).
This study includes two evaluation metrics:
the structural similarity metric (SSIM)
and peak signal-to-noise ratio (PSNR) \cite{hore:2010:ImageQualityMetrics}. 

\subsection{Trajectories}
\label{subsec:traj}
For both simulation and real acquisition, 
the acquisition sampling time and gradient raster time are both 4 $\mu$s, 
with a target matrix size of 320$\times$320.
The maximum gradient strength is 26.7 mT/m, 
and the maximum slew rate is 150 T/m/s,
\blue{which were set to limit peripheral nerve stimulation
and conform to the Nyquist criterion.}

To demonstrate the proposed model's adaptability, 
we investigated two types of initialization of waveforms: 
an undersampled in-out radial trajectory with a shorter readout time
($\sim$5 ms)
and an undersampled center-out spiral trajectory with a longer readout time
($\sim$16 ms). 
For the in-out radial initialization, the number of spokes is 16/24/32, 
and each spoke has 1280 points of acquisition (4 $\mu$s samples).
The rotation angle is equidistant between $-\pi/2$ and $\pi/2$. 
For the center-out spiral initialization, the number of spokes is 8, 
and each leaf has $\sim$4000 points of acquisition.
We used the variable-density spiral design package\footnote{\url{https://mrsrl.stanford.edu/~brian/vdspiral/}} from \cite{brian:03:vd}. 
For SPARKLING, we use 
\blue{$\tau = 0.5$ and $d=2.5$ for 16-spoke radial,
$\tau = 0.5$ and $d=2.5$ for 24-spoke radial,
$\tau = 0.6$ and $d=2.5$ for 32-spoke radial,
and $\tau = 0.5$ and $d=2$ for 8-shot spiral (\cite[Eqn.~8]{sparklingmrm}, which can also be learned as described in \cite{chaithya2021learning}.)
after grid search with CS-based reconstruction.}

\subsection{Network training and hyper-parameter setting}
\label{subsec:training}

The simulation experiments used the NYU fastMRI brain dataset to train
the trajectories and neural networks \cite{fastmri}.
The dataset consists of multiple contrasts,
including T1w (23220 slices), T2w (42250 slices), and FLAIR (5787 slices).
FastMRI's knee subset was also used
in a separate training run
to investigate the influence of training data on learned sampling patterns.
The central $320 \times 320$ region was cropped (or zero-filled).
Sensitivity maps were estimated using the ESPIRiT method \cite{espirit}
with the central 24 phase-encoding lines,
and the corresponding conjugate phase reconstruction
was regarded as the ground truth during training.

The batchsize was 4. 
The number of blocks,
or the number of outer iterations for the unrolled neural network was 6. 
The weight $\mu$ in \eqref{recon_aux} could also be learned, 
but this operation would double the computation load with minor improvement. 
We set $\mu = 2$.
The number of training epochs was set to 3
for each level of B-spline kernel length,
which is empirically enough for the training to converge.
We used $\Nlevel = 4$ optimization levels,
and $\Nepoch = 3$ % ???
so the total number of epochs was 12.
We set $\Niter = 6$ of the unrolled neural network.
For training the reconstruction network with existing trajectories
(radial, spiral, and SPARKLING-optimized), 
we also used 12 training epochs.
We used the Adam optimizer \cite{kingma:2017:AdamMethodStochastic},
with parameter $\bm{\beta}$ = $[0.5, 0.999]$,
for both trajectories \om and network parameters \thta.
The learning rate  
linearly decayed from 1e-3 to 0 for the trajectory update,
and from 1e-5 to 0 for the network update.
We did not observe obvious over-fitting phenomena on the validation set.
The training on a Intel Xeon Gold 6138 CPU and an Nvidia RTX2080Ti GPU
took around 120-150 hours%
\footnote{A demo
is available at \url{https://github.com/guanhuaw/Bjork}.
}.

\begin{figure}[htbp]
    \centerline{\includegraphics[width=0.95\columnwidth]{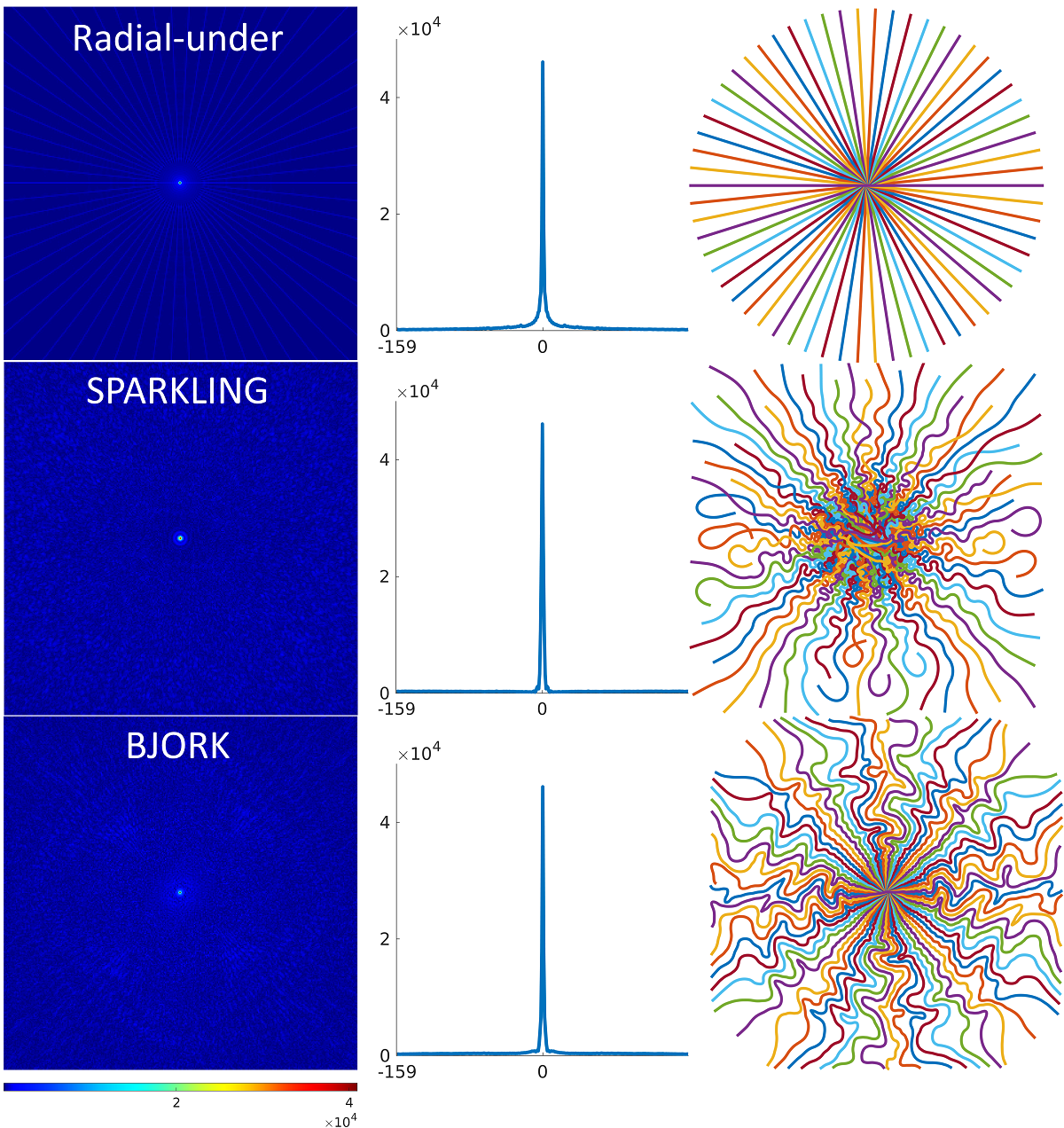}}
    \caption{PSFs of different sampling patterns.
    Each middle plot is the averaged profile of different views (angles)
    through the origin.
    The FWHM for undersampled radial, BJORK and SPARKLING is respectively 1.5, 1.6, 2.1 pixels.}
    \label{psf}
\end{figure}

\begin{figure}[htbp]
    \centerline{\includegraphics[width=0.6\columnwidth]{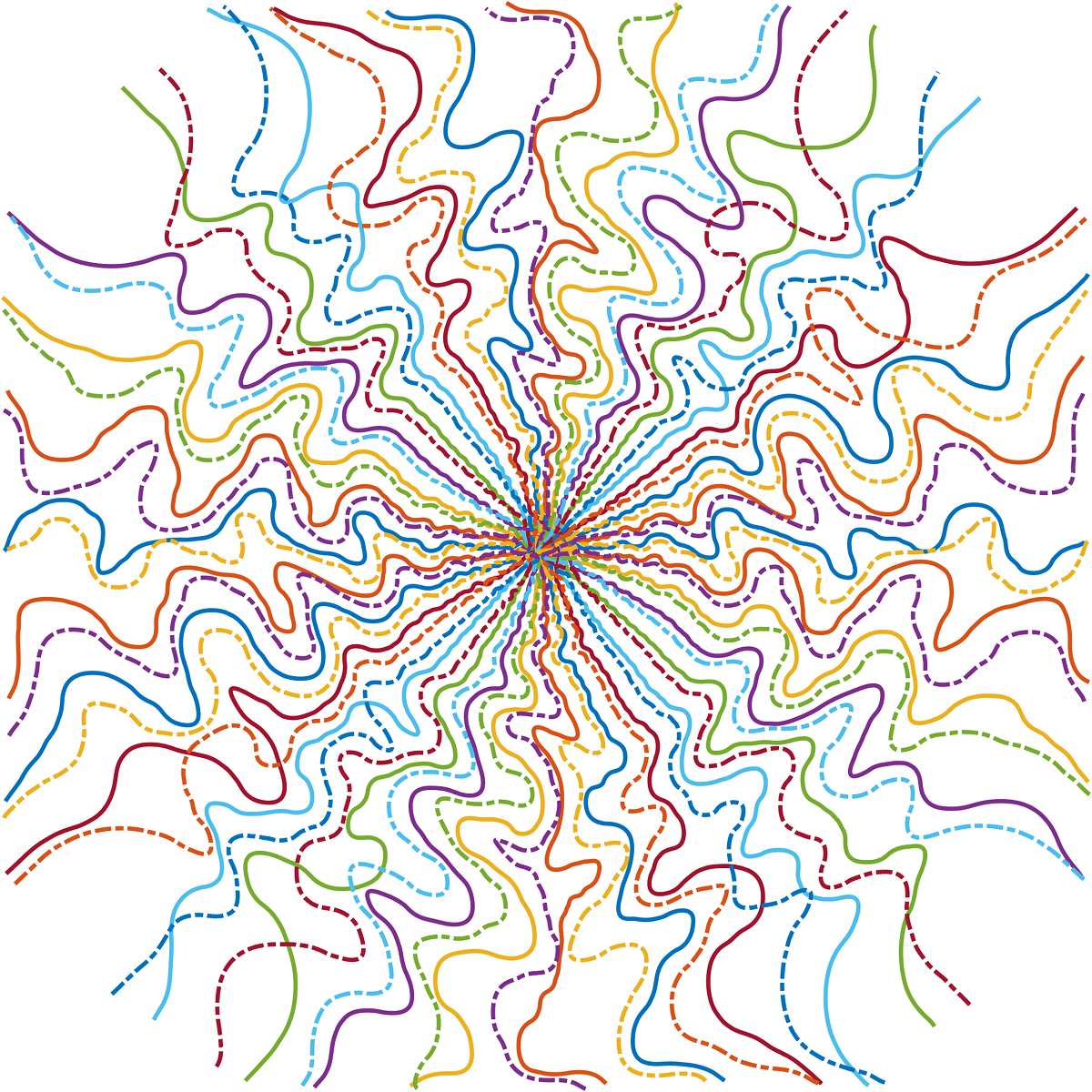}}
    \caption{The dash-dot line shows
    the $180^{\circ}$ rotated BJORK trajectory.
    The original and rotated trajectory have little overlap,
    suggesting that the BJORK automatically learned a sampling pattern
    that exploits approximate k-space Hermitian symmetry.}
    \label{conj}
\end{figure}

\begin{figure}[htbp]
    \centerline{\includegraphics[width=0.95\columnwidth]{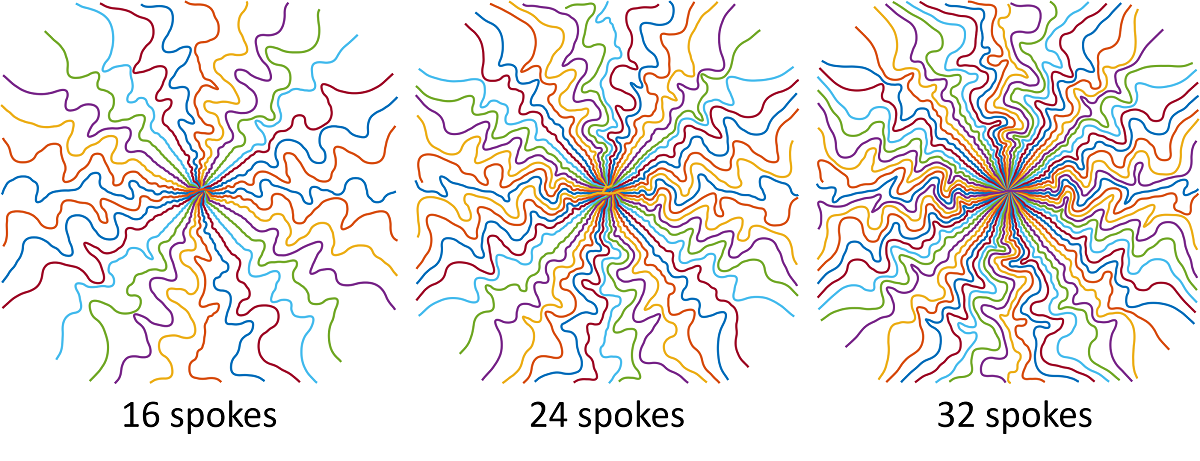}}
    \caption{Learned radial-like trajectories with different acceleration ratios.}
    \label{traj-ratios}
\end{figure}

\subsection{Prospective Studies}

\tref{protocol} details
the scanning protocols of the
RF-spoiled,
gradient echo (GRE) sequences used.
For in-vivo acquisitions,
a fat-saturation pulse was applied 
before the tip-down RF pulse.
We chose the TR and FA combination for desired T1-weighed contrast.
For radial-like sequences,
we tested a GRE sequence
with 3 different readout trajectories:
standard undersampled radial,
BJORK initialized with undersampled radial,
and SPARKLING initialized with undersampled radial.
Radial-full means the fully sampled radial trajectory.
The simulation experiments (evaluation)
and real experiments use the same readout trajectory.

We also acquired an additional dual-echo Cartesian GRE image,
for generating the sensitive map and (potentially) B0 map.
The sensitivity maps were generated
by ESPIRiT \cite{espirit} methods.
The sequences were programmed with TOPPE \cite{nielsen:2018:TOPPEFrameworkRapid},
and implemented on a GE MR750 3.0T scanner
with a Nova Medical 32 channel Rx head coil.
Subjects gave informed consent under local IRB approval.
For phantom experiments,
we used a water phantom with 3 internal cylinders.

The k-space mapping was implemented on a water phantom.
The thickness of the pencil-beam was $2$mm $\times$ $2$mm.
The trajectory estimates were based
on an average of 30 repetitions. 

%% file: t,proto.tex
\begin{table*}[htbp]
\caption{Protocols for data acquisition}
\centering
\label{protocol}
\begin{tabular}{llllllllll}
\multicolumn{10}{l}{Protocols for the prospective experiment:}   \\ \hline
Name & \textit{FOV(cm)} & \textit{dz(mm)} & \textit{Gap(mm)} & \textit{TR(ms)} & \textit{TE(ms)} & \textit{FA} & \textit{Acqs} & \textit{dt(us)} & \textit{Time} \\ \hline
Radial-like & 22*22*4 & 2 & 0.5 & 318.4 & 3.56 & 90\textdegree & 32*1280  & 4 & 0:11 \\
Radial-full & 22*22*4 & 2 & 0.5 & 318.4 & 3.56 & 90\textdegree & 320*1280 & 4 & 1:40 \\ \hline
\multicolumn{10}{l}{dz: slice thickness; Gap: gap between slices;
Acqs: number of shots * readout points; FA: flip angle} 
\end{tabular}
\end{table*}

%% file: s,res.tex
\subsubsection{Quantitative results of simulation reconstruction study}
The test set includes 1520 slices,
and the validation set includes 500 slices.
Table \ref{tab:quan} shows the quantitative results (SSIM and PSNR). 
The proposed method has significant improvement
compared with un-optimized trajectories ($P<0.005$). 
It also has improved reconstruction quality
compared with SPARKLING
considering unrolled neural network-based reconstruction. 
Compared to undersampled radial trajectory or SPARKLING trajectory,
the proposed method has a better restoration of details
and lower levels of artifacts.
In the experiment, different random seeds in training
led to very similar learned sampling trajectories.

\fref{psf} displays point spread functions
of 32-spoke radial-like trajectories.
The BJORK's PSF has a narrower central-lobe than SPARKLING
and much fewer streak artifacts than standard radial.
\fref{conj} shows the conjugate symmetry relationship
implicitly learned in the BJORK trajectory.
\fref{traj-ratios} displays optimization results under different acceleration ratios. 
\blue{\fref{simu} in the Appendix exhibits example slices.
\fref{grad_plot} in the Appendix shows the gradient waveform of one shot on one direction
(from the optimized 32-spoke radial-like trajectory)
and the corresponding slew rate.}

\input{t,quan}

\subsubsection{Multi-level optimization}
\begin{figure*}[htbp]
    \centering
    \includegraphics[width=0.95\textwidth]{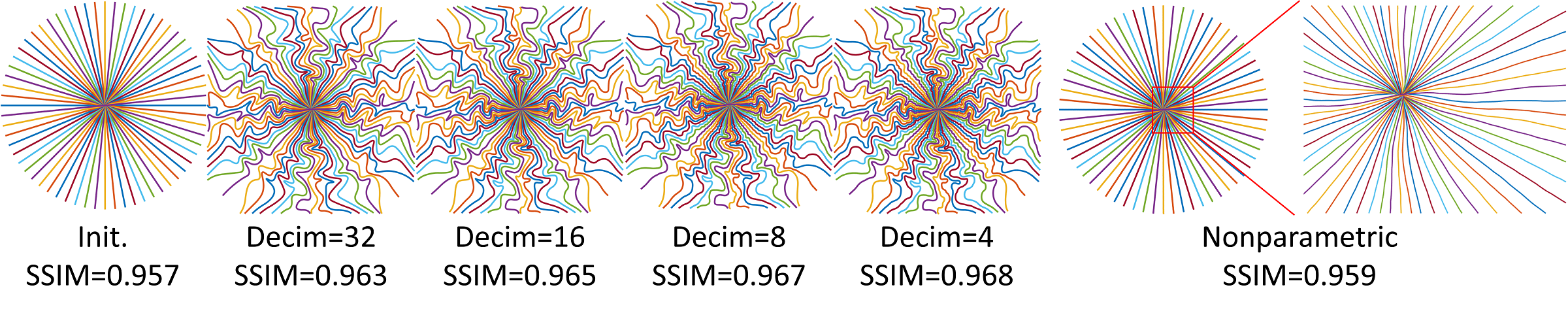}
    \caption{The evolution of the learned trajectories.
    Decim means $\Ns/L$ % why does it look bold in caption?
    in \eqref{parameterization}.
    Nonparametric means the locations of
    each sampling points are independent trainable variables,
    rather than being parameterized by quadratic B-spline kernels.
    SSIM denotes the average reconstruction quality on the evaluation set of each level.
    The rightmost zoomed-in set shows the very small perturbations
    produced by the nonparametric approach (stuck into local-minima).
    }
    \label{evo}
\end{figure*}

\fref{evo}
shows the evolution of sampling patterns
using our proposed multi-level optimization.
Different widths of the B-spline kernels introduce
different levels of improvement as the acquisition is optimized.
Also shown are the results of multi-level optimization
and a nonparametric trajectory
as used early versions of the PILOT paper \cite[versions 1-3]{pilot}.
Directly optimizing sampling points seems only to introduce
a small perturbation to the initialization. 
\blue{
\fref{loss_curve} in the Appendix shows the training losses:
the reconstruction loss $\ell(\cdot, \cdot)$,
the penalty on maximum gradient strength,
and the penalty on maximum slew rate.
}
\blue{Transitions between different B-spline kernel widths
led to a stepped training loss descent pattern.}

\subsubsection{Effect of training set}

\begin{figure}[htbp]
    \centerline{\includegraphics[width=0.75\columnwidth]{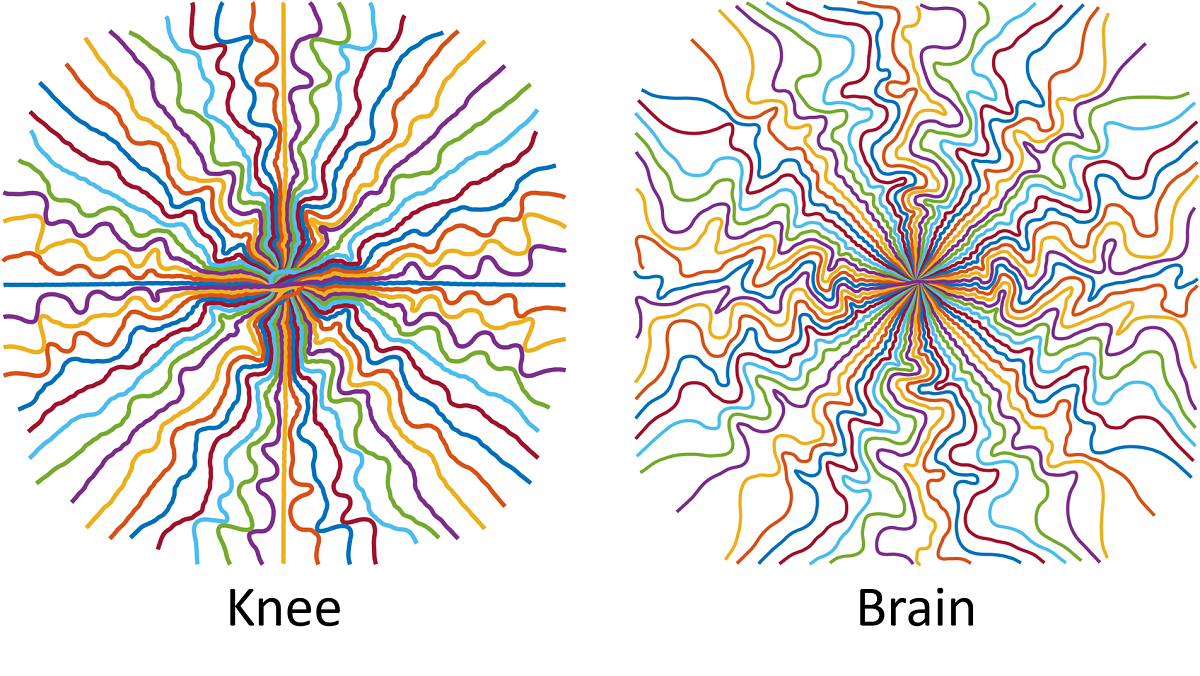}}
    \caption{Trajectories learned from different datasets.}
    \label{traj-knee-brain}
\end{figure}

\fref{traj-knee-brain} shows radial-initialized trajectories
trained by BJORK with brain and knee datasets.
Different trajectories are learned from different datasets.
We hypothesize that the difference is related to
frequency distribution of energy,
as well as the noise level,
which requires further study.
This phenomenon was also observed in \cite{bahadir:2020:DeepLearningBasedOptimizationUnderSampling}.

% To \blue{explore the influence of training set's image contrast,
% the Supplementary Materials contain a complementary experiment
% where trajectories and reconstruction networks
% are trained and tested with different contrasts.
% Within the fastMRI dataset,
% different contrasts do not influence trajectory optimization results significantly.}
% \resp{R4.6}

\subsubsection{Effect of reconstruction methods}

To test the influence of reconstruction methods on trajectory optimization,
we tried a single image-domain refinement network
as the reconstruction method in the joint learning model,
similar to PILOT's approach. 
Quadratic roughness penalty reconstruction in \eqref{init}
still is the network's input.
The initialization of the sampling pattern
is an undersampled radial trajectory.
Table \ref{reconnet}
shows that
the proposed BJORK reconstruction method (unrolled neural network, UNN)
improves reconstruction quality compared to a single end-to-end model.
Such improvements are consistent
with other comparisons between UNN methods
and image-domain CNN methods
using fixed sampling patterns (reconstruction only)
\cite{yang:2016:DeepADMMNetCompressivea,modl,schlemper:2019:SigmaNetEnsembled}.

\begin{table}[htbp]
\caption{Effect of different reconstruction networks involved in the joint learning model}
\label{reconnet}
\centering
\begin{tabular}{lll}
\hline
            & SSIM  & PSNR(dB)  \\ \hline
UNN & \textbf{0.968} & \textbf{36.9}    \\
Single U-net        & 0.934 & 32.8    \\ \hline
\end{tabular}
\end{table}

\begin{figure*}[htbp]
    \centering
    \includegraphics[width=0.95\textwidth]{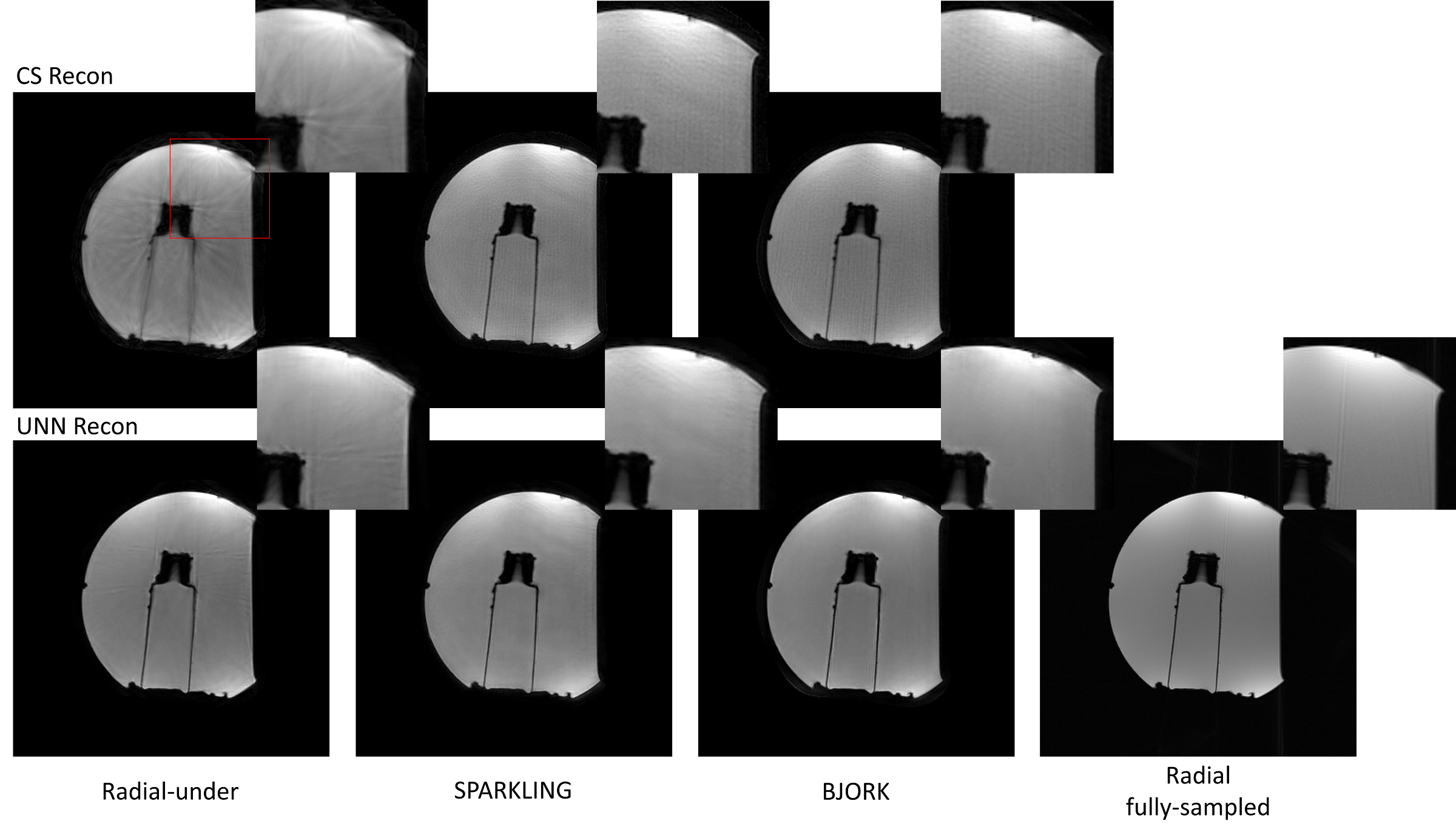}
    \caption{\blue{Representative results of the \blue{prospective} phantom experiment
    using CS-based and UNN-based reconstruction algorithms.
    The sequences involved were radial-like GRE (detailed in \tref{protocol}) with T1w contrast.
    The parameters of UNNs are trained with fastMRI dataset without fine-tuning.
    The readout length was 5.12 ms.
    The number of shots for undersampled trajectories was 32,
    and for the fully-sampled radial trajectory is 320 ($10\times$ acceleration). 
    The FOV was 22cm.
    Red boxes indicate the zoomed-in regions
    displayed 
    on the upper right corner.
    }}
    \label{phantom}
\end{figure*}

\begin{figure*}[htbp]
    \centering
    \includegraphics[width=0.98\textwidth]{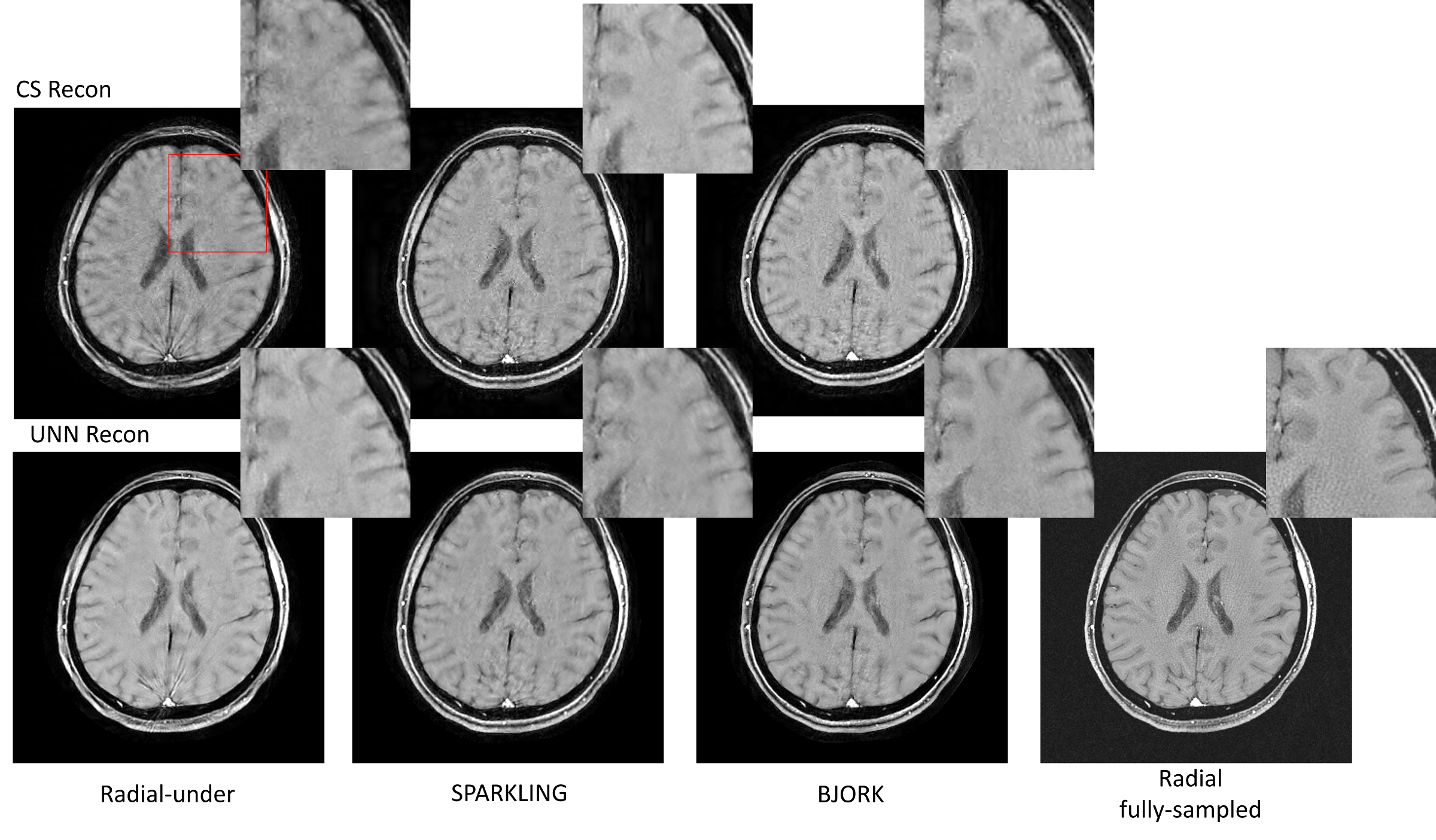}
    \caption
    {\blue{
    Results of the T1w prospective in-vivo experiment.
    The trajectories were also radial-like (detailed in \tref{protocol}).
    The parameters of UNNs are trained with
    the fastMRI dataset without fine-tuning.
    The readout time was 5.12 ms. 
    The number of shots for undersampled trajectories was 32,
    and for the fully-sampled radial trajectory is 320 ($10\times$ acceleration). 
    The FOV was 22cm.
    Red boxes indicate the zoomed-in regions
    displayed 
    on the upper right corner.
    }}
    \label{vivo}
\end{figure*}

\subsubsection{Prospective experiments}

\fref{phantom} shows the
water phantom results for different reconstruction algorithms.
The rightmost column is the fully-sampled ground truth (Radial-full).
Note that the unrolled neural network (UNN)
here was trained with fastMRI brain dataset,
and did not receive fine-tuning in all prospective experiments.
The BJORK-optimized trajectory leads to fewer artifacts
and improved contrast for the UNN-based reconstruction.

\fref{vivo} showcases one slice from the in-vivo experiment.
For CS-based reconstruction,
the undersampled radial trajectory exhibits stronger streak artifacts
than SPARKLING- and BJORK-optimized trajectories.
For UNN-based reconstruction,
all trajectories' results show reductions of
artifacts compared to CS-based reconstruction.
The proposed method restores most of the structures and fine details,
with minimal artifacts.

The Appendix also contains examples of 
reconstruction results before/after eddy currents correction,
the measurement of actual k-space trajectories,
and effectiveness of the warm initialization (quadratic least-squares reconstruction).

%% file: t,quan.tex
% \begin{table}[htbp]
% \caption{Quantitative results for simulation experiments}
% \label{tab:quan}
% \centering
% \begin{tabular}{lllll}
% \multicolumn{5}{l}{SSIM:}                                            \\ \hline
%                         &             & Standard & SPARKLING & BJORK \\ \hline
% \multirow{2}{*}{radial-like} & UNN & 0.958 & 0.963 & \textbf{0.968}   \\ 
%                         &CS  & 0.911  & \textbf{0.927} & 0.921      \\ \hline
% \multirow{2}{*}{spiral-like} & UNN &  0.985 & 0.978 & \textbf{0.989} \\
%                         & CS       &   0.958    &      0.924      &  \textbf{0.961}     \\ \hline
% \\                                               
% \end{tabular}
% \begin{tabular}{lllll}
% \multicolumn{5}{l}{PSNR (in dB):}                                            \\ \hline
% \multicolumn{1}{l}{}    &             & Standard & SPARKLING & BJORK \\ \hline
% \multirow{2}{*}{radial-like} & UNN &  35.52   &  36.31  &  \textbf{36.98}    \\
%                         & CS &   33.03    &  \textbf{34.57}    &  33.86    \\ \hline
% \multirow{2}{*}{spiral-like} & UNN & 43.86 &38.94&\textbf{44.34}     \\
%                         & CS      &  38.52    & 34.74       &  \textbf{40.85}    \\ \hline
% \end{tabular}
% \end{table}

\begin{table}[htbp]
\caption{\blue{Quantitative results for simulation experiments}}
\label{tab:quan}
\centering
\begin{tabular}{lllll}
\multicolumn{5}{l}{SSIM:}                                            \\ \hline
                        &             & Standard & SPARKLING & BJORK \\ \hline
\multirow{2}{*}{radial-like Ns=16} & UNN &  0.940 &  0.946 & \textbf{0.950}   \\ 
                        &CS  & 0.911  & 0.936 & \textbf{0.938}      \\ \hline
\multirow{2}{*}{radial-like Ns=24} & UNN & 0.950 & 0.955 & \textbf{0.959}   \\ 
                        &CS  & 0.929  & 0.943 & \textbf{0.948}      \\ \hline
\multirow{2}{*}{radial-like Ns=32} & UNN & 0.957 & 0.963 & \textbf{0.968}   \\ 
                        &CS  & 0.932  & 0.946 & \textbf{0.956}      \\ \hline
\multirow{2}{*}{spiral-like Ns=8} & UNN &  0.986 & 0.989 & \textbf{0.990} \\
                        & CS       &   0.976    &      0.978      &  \textbf{0.981}     \\ \hline
\\                                               
\end{tabular}
\begin{tabular}{lllll}
\multicolumn{5}{l}{PSNR (in dB):}                                            \\ \hline
\multicolumn{1}{l}{}    &             & Standard & SPARKLING & BJORK \\ \hline
\multirow{2}{*}{radial-like Ns=16} & UNN & 32.7    & 33.9   &  \textbf{34.3}    \\
                        & CS &  31.7     &33.6   &  \textbf{34.1}    \\ \hline
\multirow{2}{*}{radial-like Ns=24} & UNN & 34.1   & 35.0   &  \textbf{35.6}    \\
                        & CS & 33.3     & 34.6  &  \textbf{35.1}    \\ \hline
\multirow{2}{*}{radial-like Ns=32} & UNN &  35.0   &  36.0  &  \textbf{36.9}    \\
                        & CS &   33.9    &  35.7  &  \textbf{36.3}    \\ \hline
\multirow{2}{*}{spiral-like Ns=8} & UNN & 40.9 &41.7&\textbf{41.9}     \\
                        & CS      &  39.9    & 40.4       &  \textbf{40.7}    \\ \hline
\multicolumn{5}{l}{\blue{Ns: the number of shots or spokes.}} 
\end{tabular}
\end{table}

%% file: s,dis.tex
% s,dis

This paper proposes an efficient learning-based framework
for the joint design of MRI sampling trajectories and reconstruction parameters.
Defining an appropriate objective function for trajectory optimization
is an open question.
We circumvented this long-lasting problem by directly
using the reconstruction quality as the training loss function
in a supervised learning paradigm.
The workflow includes a differentiable reconstruction algorithm
for which the learning process
obtains an intermediate gradient w.r.t. the reconstruction loss.
However,
solely depending on backpropagation and stochastic gradient descent
cannot guarantee optimal results for this non-convex problem.
To improve the training effect,
we adopted several techniques,
including trajectory parameterization,
multi-level training,
warm initialization of the reconstruction network,
and an accurate approximation of NUFFT's Jacobian
\cite{wang:21:eao}.
Results show that these approaches can stabilize
the training and
%avoid sub-optimal local minima,
%or at least
provide better local minimizers than previous methods.

We trained an unrolled neural network-based reconstruction method
for non-Cartesian MRI data.
The single image-domain %end-to-end
network 
%approach
used in previous work
does not efficiently remove aliasing artifacts.
Additionally, the k-space ``hard'' data-consistency trick
for data fidelity \cite{gancs, schlemper:2018:DeepCascadeConvolutional}
is inapplicable for non-Cartesian sampling.
An unrolled algorithm can reach a balance between
data fidelity and the de-aliasing effect
across multiple iterations.
For 3D trajectory design using the proposed approach,
the unrolled method's memory consumption can be huge.
More memory-efficient reconstruction models,
such as the memory-efficient network \cite{leemput:2019:MemCNNPythonPyTorch}
should be explored in further study. 
We would also investigate
recent calibration-less unrolled neural networks,
which do not require external sensitivity maps,
and shows improved performance relative to MoDL \cite{muckley:2021:Results2020FastMRI}.

For learning-based medical imaging algorithms,
one main obstacle towards clinical application
is the gap between simulation and the physical world.
Some factors include the following.

First, inconsistency exists between the training datasets
and real-world acquisition,
such as different vendors and protocols,
posing a challenge to reconstruction algorithms' robustness and generalizability.
Our training dataset consisted of T1w/T2w/FLAIR
Fast Spin-Echo (FSE or TSE) sequences,
acquired on Siemens 1.5T/3.0T scanners.
The number of receiver channels includes 4, 8, and 16, etc.
We conducted the in-vivo/phantom experiment on a 3.0T GE scanner
equipped with a 32-channel coil.
The sequence is a GRE sequence
that has lower SNR compared to FSE sequences in the training set.
Despite the very large differences with the training set,
our work still demonstrated improved and robust results in
the in-vivo and phantom experiment,
without any fine-tuning.

We hypothesize that several factors
could contribute to the generalizability:
(1) the reconstruction network uses the quadratic 
roughness penalized reconstruction as the initialization,
normalized by the median value.
Previous works typically use the adjoint reconstruction
as the input of the network. 
In comparison, our regularized initialization helps provide
consistency between different protocols,
without too much compromise of the computation time/complexity,
(2) the PSF of the learned trajectory \resp{R4.7}
has a compact central lobe,
without significant streak artifacts.
Thus the reconstruction is basically a de-blurring/denoising task
that is a local low-level problem
and thus may require less training data
than de-aliasing problems.
For de-blurring of natural images,
networks are usually adaptive to different noise levels
and color spaces,
and require small cohorts of data
\cite{nah:2020:NTIRE2020Challenge, lugmayr:2020:NTIRE2020Challenge}.
For trajectories like radial and SPARKLING,
in contrast,
a reconstruction CNN needs to remove global aliasing artifacts,
such as the streak and ringing artifacts.
The dynamics behind the neural network's ability to
resolve such artifacts is still an unsolved question,
and the training requires a large amount of diverse data.

Secondly, it is not easy to simulate system imperfections
like eddy currents and off-resonance in the training phase.
These imperfections can greatly affect
image quality in practice.
We used a trajectory measurement method
to correct for the eddy-current effect.
Future work will incorporate field inhomogeneity 
into the workflow.

Furthermore,
even though the BJORK sampling was optimized for a UNN reconstruction method,
the results in \fref{phantom} and \fref{vivo}
suggest that
the learned trajectory is also useful
with a CS-based reconstruction method
or other model-based reconstruction algorithms.
This approach can still noticeably improve the image quality
by simply replacing the readout waveform
in the existing workflow,
promoting the applicability of the proposed approach,
similar to
\cite{bahadir:2020:DeepLearningBasedOptimizationUnderSampling}.
\resp{R1.4}
We plan to apply the general framework
to optimize a trajectory
for (convex) CS-based reconstruction
and compare to the (non-convex) open-loop UNN approach
in future work.

Though the proposed trajectory is learned via a data-driven approach,
it can also reflect the ideas behind SPARKLING and Poisson disk sampling:
sampling patterns having large gaps or tight clusters of points are inefficient,
and the sampling points should be somewhat evenly distributed
(but not too uniform).
Furthermore, BJORK appears to have learned some latent characteristics,
like the conjugate symmetry for these spin-echo training datasets.
To combine both methods' strengths,
a promising future direction is to
use SPARKLING as a primed initialization of BJORK. 

The learning used here exploited a big public data set.
As is shown in the results, knee imaging and brain imaging
led to different learned trajectories.
This demonstrates that the data set can
influence the optimization results,
as was observed in
\cite{bahadir:2020:DeepLearningBasedOptimizationUnderSampling}.
\resp{R1.4}
We also implemented a complementary experiment on a smaller training set
(results not shown).
We found that a small subset (3000 slices)
also led to similar learned trajectories.
Therefore, for some organs where a sizeable dataset
is not publicly available,
this approach may still work with small-scale private datasets.
To examine the influence of scanner models,
field strength, and sequences,
follow-up studies should investigate more diverse datasets.

The eddy-current effect poses a long-term problem
for non-Cartesian trajectories
and impedes their widespread clinical use.
This work used a simple k-space mapping
technique as the correction method.
The downside of this method is its long calibration time,
although it can be performed in a scanner's idle time.
This method is waveform-specific, 
which means that correction should be done for different trajectories.
Other methods relying on field probes can
get a more accurate correction with less time,
albeit with dedicated hardware.
In a future study, the eddy current-related artifacts
could be simulated according to the GIRF model
in the training phase,
so that the trajectory is learned to be
robust against the eddy current effect.

Aside from practical challenges with GPU memory,
the general approach described here is readily extended
from 2D to 3D sampling trajectories \cite{weiss2021optimizing}.
A more challenging future direction
is to extend the work to
dynamic imaging applications like fMRI and cardiac imaging,
where both the sampling pattern
and the reconstruction method
should exploit redundancies in the time dimension,
e.g., using low-rank models
\cite{jacob:20:slr}. 
To optimize sampling in higher dimensions,
the proposed approach should also have
additional regularization on the PNS effect.

%% file: s,supp.tex
\section{Experiments}
\subsection{Eddy-current effect}
\begin{figure}[htbp!]
    \centerline{\includegraphics[width=0.95\columnwidth]{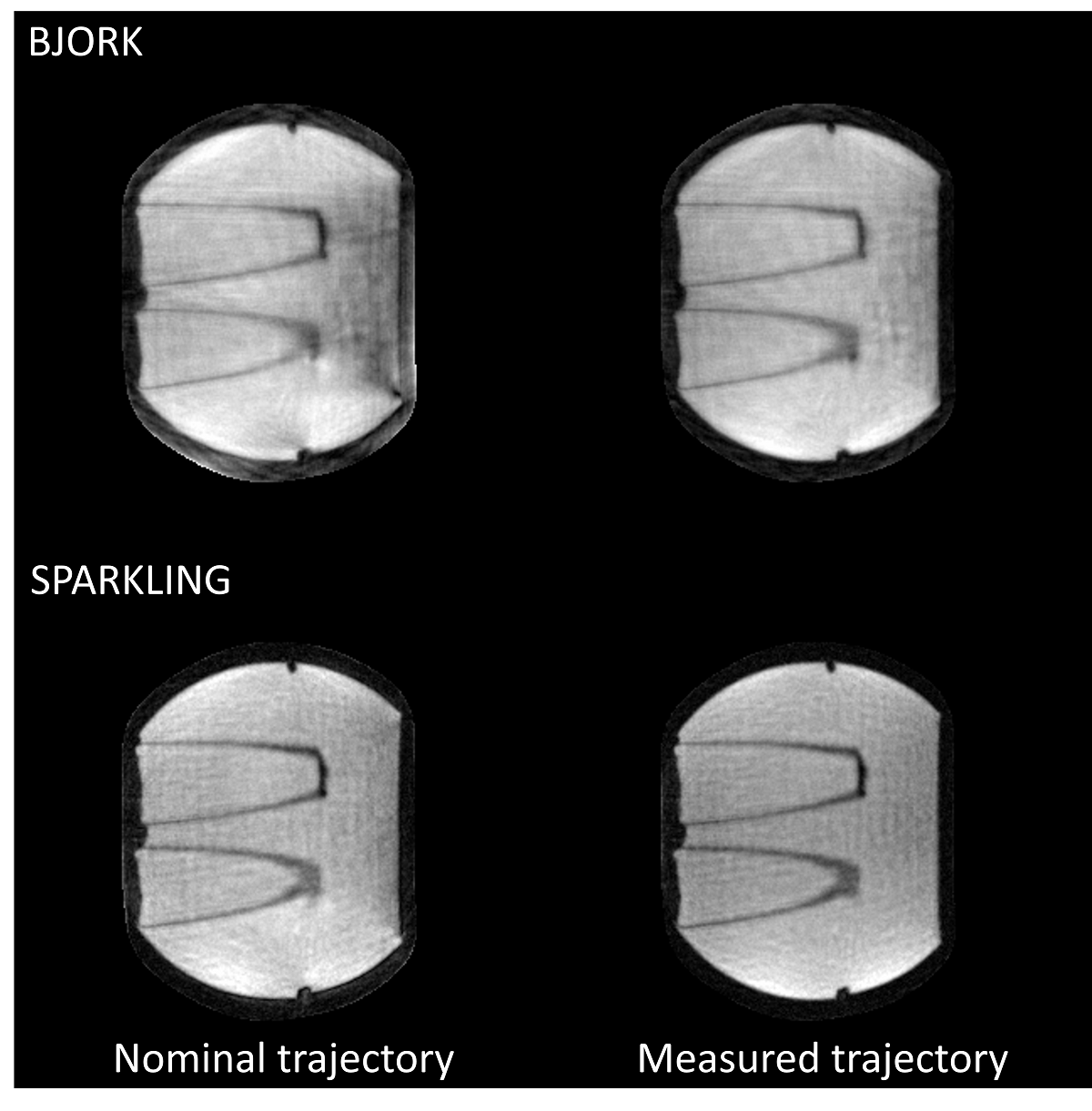}}
    \caption{Compressed sensing-based reconstruction of a water phantom.
    The left column is the reconstruction with the nominal trajectory,
    and right is with the measured trajectory.
    Reconstruction with the mapped trajectory introduces fewer artifacts.}
    \label{corr_res}
\end{figure}

\begin{figure*}[htbp!]
    \centering
    \includegraphics[width=0.98\textwidth]{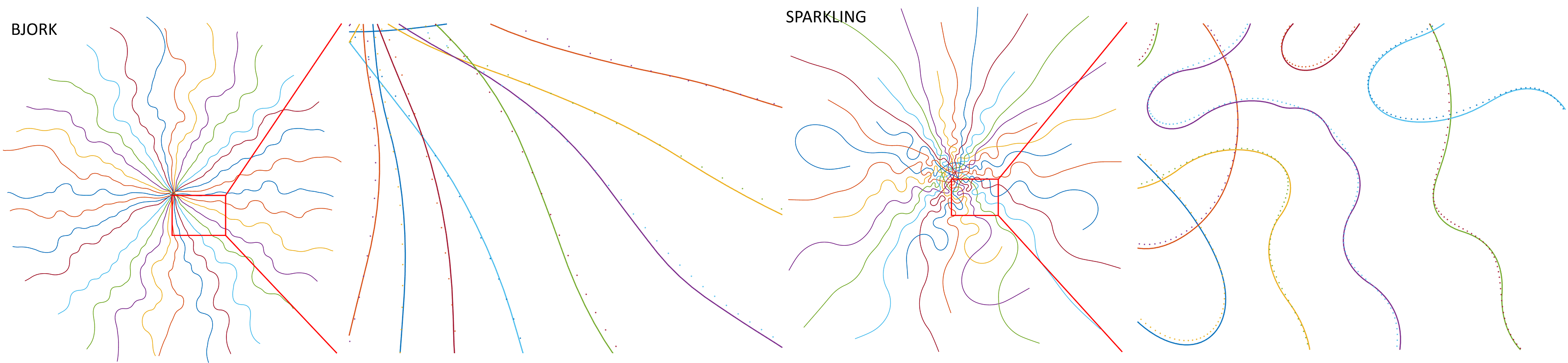}
    \caption{The measurement of the influence of the eddy currents on readout waveform.
    The solid line is the nominal trajectory, and the dotted line is the measurement.}
    \label{eddy_corr}
\end{figure*}
\fref{corr_res} displays the CS-based reconstruction of real acquisitions
reconstructed using 
both the nominally designed trajectories
and the measured trajectories.

\fref{eddy_corr} shows the results
of the trajectory measurements.
Using the measurement of the actual trajectory seems to
mitigate the influence of eddy current effects in the reconstruction results.

\begin{figure*}[htbp!]
    \centering
    \includegraphics[width=0.98\textwidth]{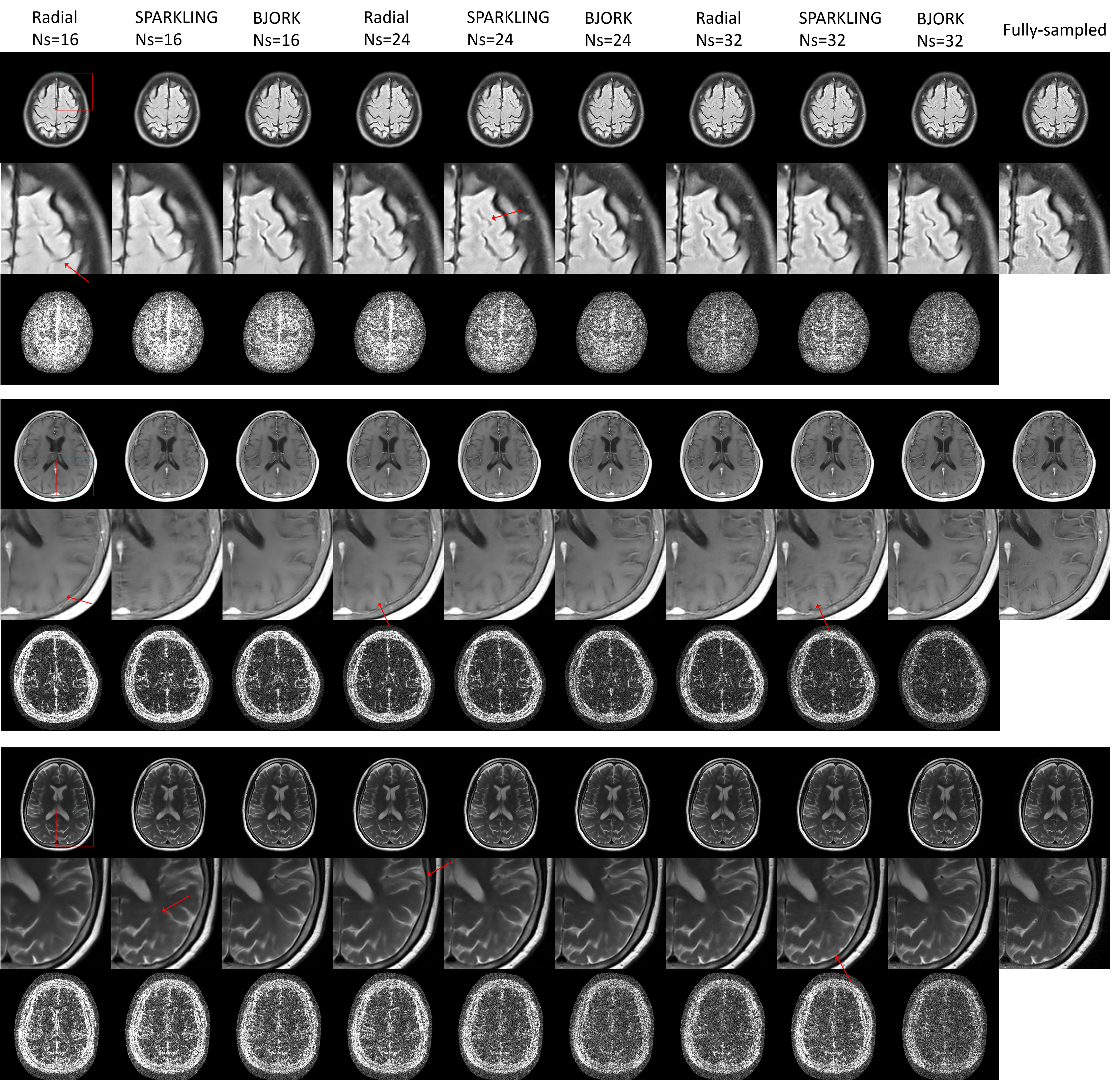}
    \caption{\blue{Examples from the simulation experiment
    using the UNN-based reconstruction algorithm,
    with three different acceleration ratios.
    \textit{Ns} stands for the number of shots or spokes.
    The first slice is FLAIR contrast.
    The second slice is T1w contrast.
    The third slice T2w contrast.
    Red boxes indicate the zoom-in region,
    and red arrows point to reconstruction artifacts/blur.
    Below the zoomed-in regions are the corresponding error maps,
    compared with fully sampled images.}
    }
    \label{simu}
\end{figure*}

\begin{figure*}[htbp!]
    \centering
    \includegraphics[width=1.8\columnwidth]{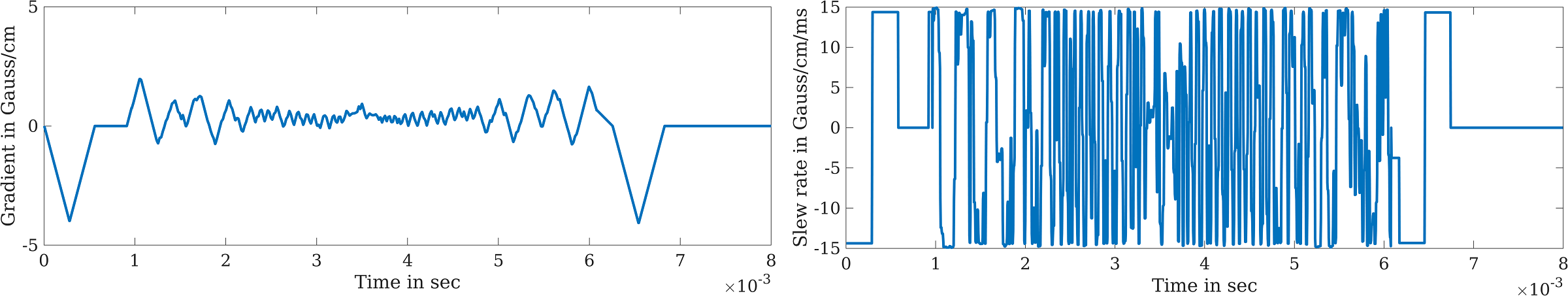}
    \caption{The gradient strength and slew rate of one spoke from 
    BJORK-optimized radial trajectory.}
    \label{grad_plot}
\end{figure*}

\begin{figure*}[htbp!]
    \centering
    \includegraphics[width=0.98\textwidth]{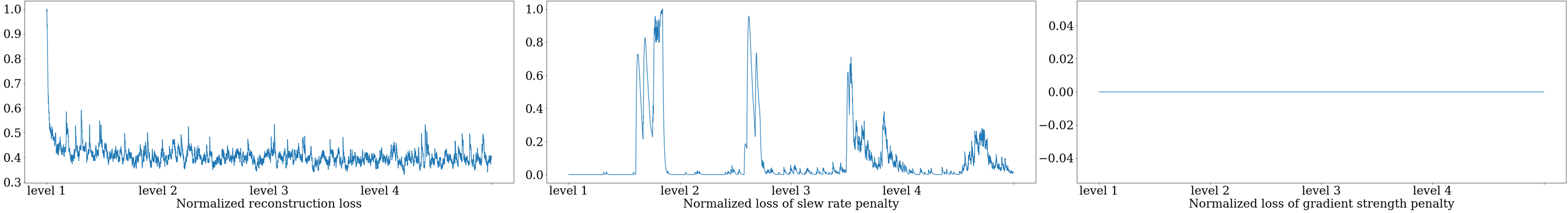}
    \caption{Smoothed training training losses of
    a 16-spoke radial-initialized sequence.
    We use 4 levels and
    each level contains 3 epochs.
    The three columns are the reconstruction loss,
    penalty on the maximum slew rate, and penalty on the maximum gradient strength.}
    \label{loss_curve}
\end{figure*}

\begin{figure}[htbp!]
    \centering
    \includegraphics[width=0.48\textwidth]{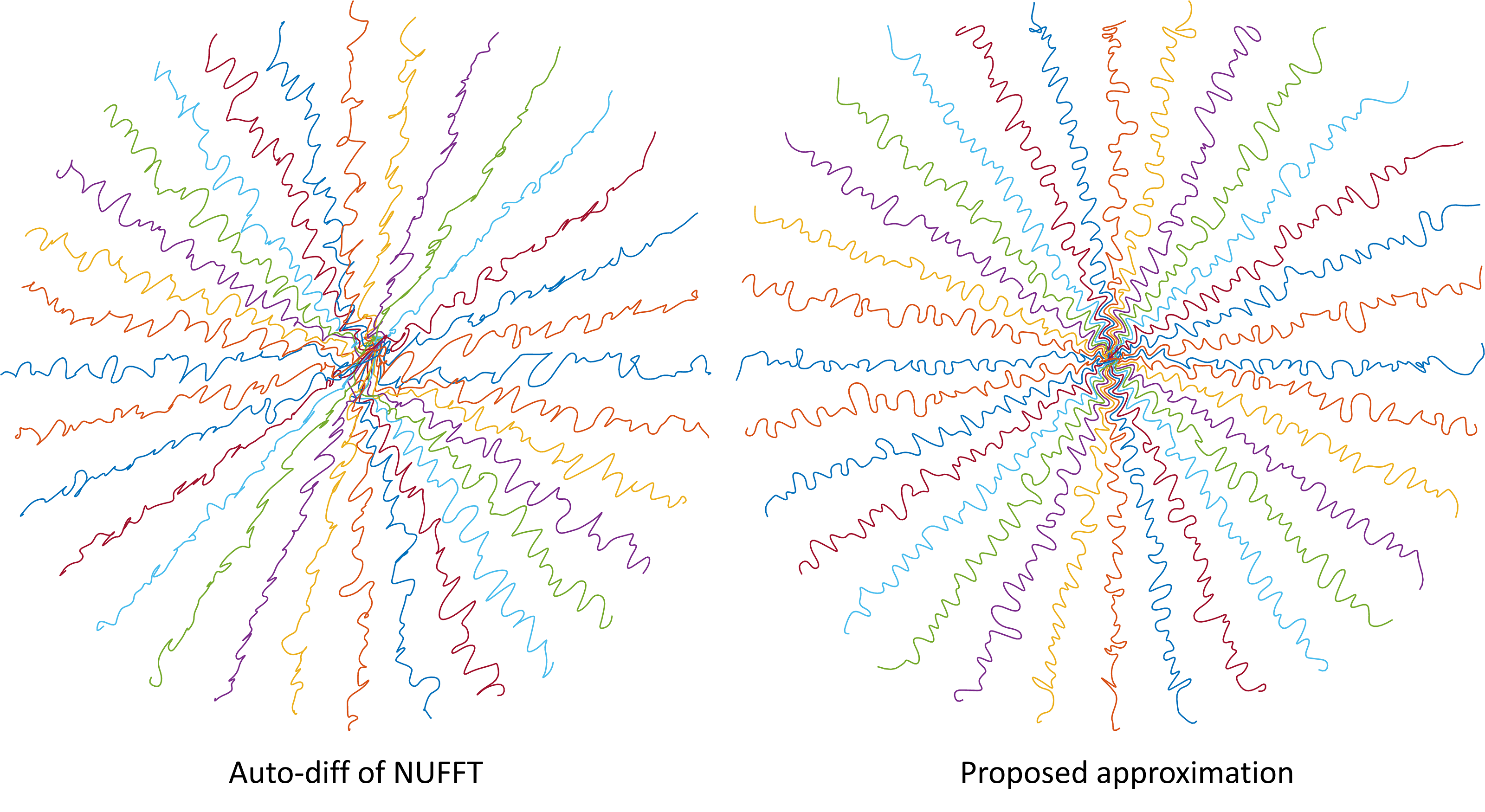}
    \caption{The learned trajectories with descent directions
    calculated by different methods.}
    \label{traj}
\end{figure}

% \subsection{Simulation Experiments}

% \fref{simu} shows examples of the UNN-based reconstruction
% for different trajectories in the simulation experiment. 
% Compared with the undersampled radial trajectory
% or the SPARKLING trajectory,
% the proposed BJORK trajectory leads to
% fewer artifacts and 
% better restoration of fine structures.

\subsection{Cross contrast validation}

In this experiment,
we trained the model with
one image contrast from the fastMRI brain dataset
(without simulated additive noise),
and tested the learned trajectory with all contrasts
(with simulated additive Gaussian noise
whose variance is $10^{-3}$ of the mean magnitude of the signal).
Each contrast has 4500 training slices and 500 test slices.
We fine-tuned the reconstruction unrolled neural network
for different test contrasts.
The initialization is a 16-spoke radial trajectory.
\tref{t-cross} shows the average reconstruction quality.
The learned trajectories are insensitive to different contrasts within the 
fastMRI dataset.

\resp{R4.6}

\begin{table}[htbp!]
\caption{Effect of different contrasts on learned models.}
\centering
\begin{tabular}{llll}
\hline
\backslashbox{test}{training} & T1w & T2w & FLAIR \\ \hline
T1w+noise   & 0.981    &0.980     & 0.981      \\
T2w+noise   & 0.951    & 0.953    & 0.953      \\
FLAIR+noise & 0.974    &0.974     & 0.975       \\ \hline
\end{tabular}
\label{t-cross}
\end{table}

\subsection{Accurate Jacobian of NUFFT}

We compared the trajectories learned with 
different NUFFT Jacobian calculation methods:
our accurate DFT approximation methods
\cite{wang:21:eao},
and using auto-differentiation of NUFFT
(the approach used in PILOT \cite{pilot}).
To save time,
we used only one level of parameterization  ($Decim.$ = 4)
and 6 epochs.
In \fref{traj},
our approximation method leads to a learned trajectory
consistent with intuition:
sampling points should not be clustered
or too distant from each other.
The quantitative reconstruction results also demonstrate
significant improvement (950 test slices, SSIM: 0.930 vs. 0.957.)

\subsection{Benefit of the warm initialization}
We compared two inputs of the unrolled neural network:
the adjoint of undersampling signal ($\A' \y$)
and quadratic roughness penalized reconstruction
$(\A'\A+ \lambda \R'\R)^{-1}\A'\y$.
The experiment optimized a 16-spoke radial trajectory and used 1520 test slices.
The average reconstruction quality (SSIM values) of the 
two settings are 0.944 and 0.950, respectively.
% The warm initialization would significantly improve 
% the reconstruction quality.

%% file: root.bbl
\begin{thebibliography}{10}
\providecommand{\url}[1]{#1}
\csname url@samestyle\endcsname
\providecommand{\newblock}{\relax}
\providecommand{\bibinfo}[2]{#2}
\providecommand{\BIBentrySTDinterwordspacing}{\spaceskip=0pt\relax}
\providecommand{\BIBentryALTinterwordstretchfactor}{4}
\providecommand{\BIBentryALTinterwordspacing}{\spaceskip=\fontdimen2\font plus
\BIBentryALTinterwordstretchfactor\fontdimen3\font minus
  \fontdimen4\font\relax}
\providecommand{\BIBforeignlanguage}[2]{{%
\expandafter\ifx\csname l@#1\endcsname\relax
\typeout{** WARNING: IEEEtran.bst: No hyphenation pattern has been}%
\typeout{** loaded for the language `#1'. Using the pattern for}%
\typeout{** the default language instead.}%
\else
\language=\csname l@#1\endcsname
\fi
#2}}
\providecommand{\BIBdecl}{\relax}
\BIBdecl

\bibitem{lauterbur1973image}
P.~C. Lauterbur, ``Image formation by induced local interactions: examples
  employing nuclear magnetic resonance,'' \emph{Nature}, vol. 242, no. 5394,
  pp. 190--191, 1973.

\bibitem{ahn1986spiral}
C.~B. Ahn, J.~H. Kim, and Z.~H. Cho, ``High-{{speed spiral}}-{{scan echo planar
  NMR imaging}}-{{I}},'' \emph{IEEE Trans. Med. Imag.}, vol.~5, no.~1, pp.
  2--7, Mar. 1986.

\bibitem{larkman:2007:ParallelMagneticResonance}
D.~J. Larkman and R.~G. Nunes, ``Parallel magnetic resonance imaging,''
  \emph{Phys. Med. Biol.}, vol.~52, no.~7, pp. R15--R55, Mar. 2007.

\bibitem{wang:2010:VariableDensityCompressed}
Z.~Wang and G.~R. Arce, ``Variable {{density compressed image sampling}},''
  \emph{IEEE Trans. Image Proc.}, vol.~19, no.~1, pp. 264--270, Jan. 2010.

\bibitem{knoll:2011:AdaptedRandomSampling}
F.~Knoll, C.~Clason, C.~Diwoky, and R.~Stollberger, ``Adapted random sampling
  patterns for accelerated {{MRI}},'' \emph{Magn. Reson. Mater. Phys. Biol.
  Med.}, vol.~24, no.~1, pp. 43--50, Feb. 2011.

\bibitem{seeger:2010:OptimizationKspaceTrajectories}
M.~Seeger, H.~Nickisch, R.~Pohmann, and B.~Sch{\"o}lkopf, ``Optimization of
  {k-space} trajectories for compressed sensing by {{Bayesian}} experimental
  design,'' \emph{Magn. Reson. Med.}, vol.~63, no.~1, pp. 116--126, 2010.

\bibitem{chauffert:2013:VariableDensityCompressed}
N.~Chauffert, P.~Ciuciu, and P.~Weiss, ``Variable density compressed sensing in
  {{MRI}}. {{Theoretical}} vs heuristic sampling strategies,'' in \emph{2013
  {{IEEE}} 10th {{Intl. Symp.}} on {{Biomed. Imag.}} (ISBI)}, Apr. 2013, pp.
  298--301.

\bibitem{robison:2019:CorrectionB0Eddy}
R.~K. Robison, Z.~Li, D.~Wang, M.~B. Ooi, and J.~G. Pipe, ``Correction of
  {{B0}} eddy current effects in spiral {{MRI}},'' \emph{Magn. Reson. Med.},
  vol.~81, no.~4, pp. 2501--2513, 2019.

\bibitem{lee:2003:Fast3DImaging}
J.~H. Lee, B.~A. Hargreaves, B.~S. Hu, and D.~G. Nishimura, ``Fast {{3D}}
  imaging using variable-density spiral trajectories with applications to limb
  perfusion,'' \emph{Magn. Reson. Med.}, vol.~50, no.~6, pp. 1276--1285, 2003.

\bibitem{winkelmann:2007:OptimalRadialProfile}
S.~Winkelmann, T.~Schaeffter, T.~Koehler, H.~Eggers, and O.~Doessel, ``An
  {{optimal radial profile order based}} on the {{golden ratio}} for
  {{time}}-{{resolved MRI}},'' \emph{IEEE Trans. Med. Imag.}, vol.~26, no.~1,
  pp. 68--76, Jan. 2007.

\bibitem{bilgic:2015:WaveCAIPIHighlyAccelerated}
B.~Bilgic \emph{et~al.}, ``Wave-{{CAIPI}} for highly accelerated {{3D}}
  imaging,'' \emph{Magn. Reson. Med.}, vol.~73, no.~6, pp. 2152--2162, 2015.

\bibitem{bilgin:08:random}
A.~Bilgin, T.~Troouard, A.~Gmitro, and M.~Altbach, ``Randomly perturbed radial
  trajectories for compressed sensing {MRI},'' in \emph{Proc. Intl. Soc. Magn.
  Reson. Med. (ISMRM)}, vol.~16, 2008, p. 3152.

\bibitem{lustig:2008:FastMethodDesigning}
M.~Lustig, S.~Kim, and J.~M. Pauly, ``A fast method for designing time-optimal
  gradient waveforms for arbitrary {k-space} trajectories,'' \emph{IEEE Trans.
  Med. Imag.}, vol.~27, no.~6, pp. 866--873, Jun. 2008.

\bibitem{sabat:2003:ThreeDimensionalKspace}
S.~Sabat, R.~Mir, M.~Guarini, A.~Guesalaga, and P.~Irarrazaval, ``Three
  dimensional {k-space} trajectory design using genetic algorithms,''
  \emph{Magn. Reson. Imag.}, vol.~21, no.~7, pp. 755--764, Sep. 2003.

\bibitem{sparklingmrm}
C.~Lazarus \emph{et~al.}, ``{SPARKLING:} variable-density k-space filling
  curves for accelerated {T2*-weighted} {MRI},'' \emph{{Mag. Res. Med.}},
  vol.~81, no.~6, pp. {3643--61}, Jun. 2019.

\bibitem{weiss2021optimizing}
\BIBentryALTinterwordspacing
P.~Weiss, A.~Massire, A.~Vignaud, P.~Ciuciu \emph{et~al.}, ``{Optimizing full
  3D SPARKLING trajectories for high-resolution T2*-weighted Magnetic Resonance
  Imaging},'' 2021. [Online]. Available: \url{http://arxiv.org/abs/2108.02991}
\BIBentrySTDinterwordspacing

\bibitem{lazarus20203d}
C.~Lazarus \emph{et~al.}, ``{3D variable-density SPARKLING trajectories for
  high-resolution T2*-weighted magnetic resonance imaging},'' \emph{NMR Bio.},
  vol.~33, no.~9, p. e4349, 2020.

\bibitem{fong:2009:BlackboxFastMultipole}
W.~Fong and E.~Darve, ``The black-box fast multipole method,'' \emph{J. Comput.
  Phys.}, vol. 228, no.~23, pp. 8712--8725, Dec. 2009.

\bibitem{chauffert:2016:ProjectionAlgorithmGradient}
N.~Chauffert, P.~Weiss, J.~Kahn, and P.~Ciuciu, ``A {{projection algorithm}}
  for {{gradient waveforms design}} in {{magnetic resonance imaging}},''
  \emph{IEEE Trans. Med. Imag.}, vol.~35, no.~9, pp. 2026--2039, Sep. 2016.

\bibitem{elmalem:2018:LearnedPhaseCoded}
S.~Elmalem, R.~Giryes, and E.~Marom, ``Learned phase coded aperture for the
  benefit of depth of field extension,'' \emph{Opt. Exp.}, vol.~26, no.~12, pp.
  15\,316--15\,331, Jun. 2018.

\bibitem{huijben:2020:LearningSubSamplingSignal}
I.~A.~M. Huijben, B.~S. Veeling, K.~Janse, M.~Mischi, and R.~J.~G. van Sloun,
  ``Learning {{sub}}-{{sampling}} and {{signal recovery with applications}} in
  {{ultrasound imaging}},'' \emph{IEEE Trans. Med. Imag.}, vol.~39, no.~12, pp.
  3955--3966, Dec. 2020.

\bibitem{bahadir:2020:DeepLearningBasedOptimizationUnderSampling}
C.~D. Bahadir, A.~Q. Wang, A.~V. Dalca, and M.~R. Sabuncu,
  ``Deep-{{learning}}-{{based optimization}} of the {{under}}-{{sampling
  pattern}} in {{MRI}},'' \emph{IEEE Trans. Comput. Imag.}, vol.~6, pp.
  1139--1152, 2020.

\bibitem{huijben:2020:LearningSamplingModelBased}
I.~A.~M. Huijben, B.~S. Veeling, and R.~J.~G. van Sloun, ``Learning
  {{sampling}} and {{model}}-{{based signal recovery}} for {{compressed sensing
  MRI}},'' in \emph{2020 {{IEEE Intl. Conf.}} on {{Acous.}}, {{Speech}} and
  {{Sig. Proc.}} ({{ICASSP}})}, May 2020, pp. 8906--8910.

\bibitem{sanchez:2020:ScalableLearningBasedSampling}
T.~Sanchez \emph{et~al.}, ``Scalable {{learning}}-{{based sampling
  optimization}} for {{compressive dynamic MRI}},'' in \emph{2020 {{IEEE Intl.
  Conf.}} on {{Acous.}}, {{Speech}} and {{Sig. Proc.}} ({{ICASSP}})}, May 2020,
  pp. 8584--8588.

\bibitem{zibetti2020fast}
M.~V.~W. Zibetti, G.~T. Herman, and R.~R. Regatte, ``Fast data-driven learning
  of parallel {{MRI}} sampling patterns for large scale problems,'' \emph{Sci
  Rep}, vol.~11, no.~1, p. 19312, Sep. 2021.

\bibitem{qian:2015:SubsetSelectionPareto}
C.~Qian, Y.~Yu, and Z.-H. Zhou, ``Subset selection by {{Pareto}}
  optimization,'' in \emph{Proc. Intl. Conf. Neur. Info. Proc. Sys. (NeurIPS)},
  ser. {{NIPS}}'15, Dec. 2015, pp. 1774--1782.

\bibitem{jin:2019:SelfSupervisedDeepActive}
\BIBentryALTinterwordspacing
K.~H. Jin, M.~Unser, and K.~M. Yi, ``Self-{{supervised deep active accelerated
  MRI}},'' 2020. [Online]. Available: \url{http://arxiv.org/abs/1901.04547}
\BIBentrySTDinterwordspacing

\bibitem{rl:david}
D.~Zeng, C.~Sandino, D.~Nishimura, S.~Vasanawala, and J.~Cheng, ``Reinforcement
  learning for online undersampling pattern optimization,'' in \emph{Proc.
  Intl. Soc. Magn. Reson. Med. (ISMRM)}, 2019, p. 1092.

\bibitem{sherry:20:lts}
F.~Sherry, M.~Benning, J.~C. D.~. Reyes, M.~J. Graves, G.~Maierhofer,
  G.~Williams, C.-B. {Schonlieb}, and M.~J. Ehrhardt, ``Learning the sampling
  pattern for {MRI},'' \emph{{IEEE Trans. Med. Imag.}}, vol.~39, no.~12, pp.
  {4310--21}, Dec. 2020.

\bibitem{aggarwal:2020:JointOptimizationSampling}
H.~K. Aggarwal and M.~Jacob, ``Joint {{optimization}} of {{sampling patterns}}
  and {{deep priors}} for {{improved parallel MRI}},'' in \emph{2020 {{IEEE
  Intl. Conf.}} on {{Acous.}}, {{Speech}} and {{Sig. Proc.}} ({{ICASSP}})}, May
  2020, pp. 8901--8905.

\bibitem{pilot}
T.~Weiss, O.~Senouf, S.~Vedula, O.~Michailovich, M.~Zibulevsky, and
  A.~Bronstein, ``{PILOT: Physics-Informed Learned Optimized Trajectories for
  Accelerated MRI},'' \emph{MELBA}, pp. 1--23, 2021.

\bibitem{unet}
O.~Ronneberger, P.~Fischer, and T.~Brox, ``U-{{Net}}: {{convolutional
  networks}} for {{biomedical image segmentation}},'' in \emph{Med. {{Imag.
  Comput.}} and {{Comput.}}-{{Assist. Interv.}} ({{MICCAI}})}, 2015, pp.
  234--241.

\bibitem{wang:21:eao}
\BIBentryALTinterwordspacing
G.~Wang and J.~A. Fessler, ``Efficient approximation of {Jacobian} matrices
  involving a non-uniform fast {Fourier} transform {(NUFFT)},'' 2021. [Online].
  Available: \url{https://arxiv.org/abs/2111.02912}
\BIBentrySTDinterwordspacing

\bibitem{modl}
H.~K. Aggarwal, M.~P. Mani, and M.~Jacob, ``{MoDL:} model-based deep learning
  architecture for inverse problems,'' \emph{{IEEE Trans. Med. Imag.}},
  vol.~38, no.~2, pp. {394--405}, Feb. 2019.

\bibitem{pruessmann:2001:AdvancesSensitivityEncoding}
K.~P. Pruessmann, M.~Weiger, P.~B{\"o}rnert, and P.~Boesiger, ``Advances in
  sensitivity encoding with arbitrary {k-space} trajectories,'' \emph{Magn.
  Reson. Med.}, vol.~46, no.~4, pp. 638--651, 2001.

\bibitem{hao:2016:JointDesignExcitationa}
S.~Hao, J.~A. Fessler, D.~C. Noll, and J.-F. Nielsen, ``Joint {{design}} of
  {{excitation}} k-{{space trajectory}} and {{RF pulse}} for {{small}}-{{tip 3D
  tailored excitation}} in {{MRI}},'' \emph{IEEE Trans. Med. Imag.}, vol.~35,
  no.~2, pp. 468--479, Feb. 2016.

\bibitem{Nielsen:16:Improved}
F.~J. Nielsen~JF, Sun~H and N.~DC, ``Improved gradient waveforms for small-tip
  {3D} spatially tailored excitation using iterated local search,'' in
  \emph{Proc. Intl. Soc. Magn. Reson. Med. (ISMRM)}, 2016, p. 1013.

\bibitem{boyer2016generation}
C.~Boyer, N.~Chauffert, P.~Ciuciu, J.~Kahn, and P.~Weiss, ``On the generation
  of sampling schemes for magnetic resonance imaging,'' \emph{SIAM J. Imag.
  Sci.}, vol.~9, no.~4, pp. 2039--2072, 2016.

\bibitem{yang:2016:DeepADMMNetCompressivea}
Y.~Yang, J.~Sun, H.~Li, and Z.~Xu, ``Deep {{ADMM}}-{{Net}} for compressive
  sensing {{MRI}},'' in \emph{Proc. of the 30th {{Intl. Conf.}} on {{Neur.
  Info. Proc. Sys.}} (NIPS)}, Dec. 2016, pp. 10--18.

\bibitem{hammernik:2018:LearningVariationalNetwork}
K.~Hammernik \emph{et~al.}, ``Learning a variational network for reconstruction
  of accelerated {{MRI}} data,'' \emph{Magn. Reson. Med.}, vol.~79, no.~6, pp.
  3055--3071, 2018.

\bibitem{schlemper:2019:SigmaNetEnsembled}
\BIBentryALTinterwordspacing
J.~Schlemper, C.~Qin, J.~Duan, R.~M. Summers, and K.~Hammernik, ``{Sigma}-net:
  {{ensembled iterative deep neural networks}} for {{accelerated parallel MR
  image reconstruction}},'' 2020. [Online]. Available:
  \url{http://arxiv.org/abs/1912.05480}
\BIBentrySTDinterwordspacing

\bibitem{fessler:05:tbi}
J.~A. Fessler, S.~Lee, V.~T. Olafsson, H.~R. Shi, and D.~C. Noll,
  ``Toeplitz-based iterative image reconstruction for {MRI} with correction for
  magnetic field inhomogeneity,'' \emph{{IEEE Trans. Sig. Proc.}}, vol.~53,
  no.~9, pp. {3393--402}, Sep. 2005.

\bibitem{muckley:20:tah}
M.~J. Muckley, R.~Stern, T.~Murrell, and F.~Knoll, ``{TorchKbNufft}: A
  high-level, hardware-agnostic non-uniform fast fourier transform,'' in
  \emph{ISMRM Workshop on Data Sampling \& Image Reconstruction}, 2020.

\bibitem{gregor:2010:LearningFastApproximations}
K.~Gregor and Y.~LeCun, ``Learning fast approximations of sparse coding,'' in
  \emph{Proc. of the 27th {{Intl. Conf.}} on {{Mach. Learn.}} (ICML)}, Jun.
  2010, pp. 399--406.

\bibitem{DIDN}
S.~Yu, B.~Park, and J.~Jeong, ``Deep iterative down-up {CNN} for image
  denoising,'' in \emph{Proc. of the IEEE Conf. on Comput. Vis. and Patt.
  Recog. Work. (CVPRW)}, 2019, pp. 0--0.

\bibitem{vannesjo:2013:GradientSystemCharacterization}
S.~J. Vannesjo \emph{et~al.}, ``Gradient system characterization by impulse
  response measurements with a dynamic field camera,'' \emph{Magn. Reson.
  Med.}, vol.~69, no.~2, pp. 583--593, 2013.

\bibitem{duyn:1998:SimpleCorrectionMethod}
J.~H. Duyn, Y.~Yang, J.~A. Frank, and J.~W. {van der Veen}, ``Simple correction
  method for {k-space} trajectory deviations in {{MRI}},'' \emph{J. Magn.
  Reson.}, vol. 132, pp. 150--153, May 1998.

\bibitem{nielsen:2018:TOPPEFrameworkRapid}
J.-F. Nielsen and D.~C. Noll, ``{{TOPPE}}: {{a}} framework for rapid
  prototyping of {{MR}} pulse sequences,'' \emph{Magn. Reson. Med.}, vol.~79,
  no.~6, pp. 3128--3134, 2018.

\bibitem{hore:2010:ImageQualityMetrics}
A.~Hor{\'e} and D.~Ziou, ``Image {{quality metrics}}: {{PSNR}} vs. {{SSIM}},''
  in \emph{{{Intl. Conf.}} on {{Patn. Recog.}} (ICPR)}, Aug. 2010, pp.
  2366--2369.

\bibitem{brian:03:vd}
J.~H. Lee, B.~A. Hargreaves, B.~S. Hu, and D.~G. Nishimura, ``Fast {3D} imaging
  using variable-density spiral trajectories with applications to limb
  perfusion,'' \emph{Magn. Reson. Med.}, vol.~50, no.~6, pp. 1276--1285, 2003.

\bibitem{chaithya2021learning}
G.~Chaithya, Z.~Ramzi, and P.~Ciuciu, ``Learning the sampling density in {2D
  SPARKLING MRI} acquisition for optimized image reconstruction,'' in
  \emph{2021 29th Euro. Sig. Proc. Conf. (EUSIPCO)}.\hskip 1em plus 0.5em minus
  0.4em\relax IEEE, 2021, pp. 960--964.

\bibitem{fastmri}
\BIBentryALTinterwordspacing
J.~Zbontar \emph{et~al.}, ``{fastMRI: An} open dataset and benchmarks for
  accelerated {MRI},'' 2018. [Online]. Available:
  \url{http://arxiv.org/abs/1811.08839}
\BIBentrySTDinterwordspacing

\bibitem{espirit}
M.~Uecker \emph{et~al.}, ``{ESPIRiT - an} eigenvalue approach to
  autocalibrating parallel {MRI:} {where} {SENSE} meets {GRAPPA},'' \emph{{Mag.
  Reson. Med.}}, vol.~71, no.~3, pp. {990--1001}, Mar. 2014.

\bibitem{kingma:2017:AdamMethodStochastic}
\BIBentryALTinterwordspacing
D.~P. Kingma and J.~Ba, ``Adam: {{a method}} for {{stochastic optimization}},''
  2017. [Online]. Available: \url{http://arxiv.org/abs/1412.6980}
\BIBentrySTDinterwordspacing

\bibitem{gancs}
M.~Mardani \emph{et~al.}, ``Deep generative adversarial neural networks for
  compressive sensing {MRI},'' \emph{IEEE Trans. Med. Imag.}, vol.~38, no.~1,
  pp. 167--179, 2018.

\bibitem{schlemper:2018:DeepCascadeConvolutional}
J.~Schlemper, J.~Caballero, J.~V. Hajnal, A.~N. Price, and D.~Rueckert, ``A
  {{deep cascade}} of {{convolutional neural networks}} for {{dynamic MR image
  reconstruction}},'' \emph{IEEE Trans. Med. Imag.}, vol.~37, no.~2, pp.
  491--503, Feb. 2018.

\bibitem{leemput:2019:MemCNNPythonPyTorch}
S.~C. van~de Leemput, J.~Teuwen, B.~van Ginneken, and R.~Manniesing,
  ``{{MemCNN}}: {{a Python}}/{{PyTorch}} package for creating memory-efficient
  invertible neural networks,'' \emph{J. Open Source Softw.}, vol.~4, no.~39,
  p. 1576, Jul. 2019.

\bibitem{muckley:2021:Results2020FastMRI}
M.~J. Muckley \emph{et~al.}, ``Results of the 2020 {fastMRI} challenge for
  machine learning {MR} image reconstruction,'' \emph{IEEE Trans. Med.
  Imaging}, vol.~40, no.~9, pp. 2306--2317, Sep. 2021.

\bibitem{nah:2020:NTIRE2020Challenge}
S.~Nah \emph{et~al.}, ``{{NTIRE}} 2020 {{challenge}} on {{image}} and {{video
  deblurring}},'' in \emph{Proc. of the IEEE Conf. on Comput. Vis. and Patt.
  Recog. Work. (CVPRW)}, 2020, pp. 416--417.

\bibitem{lugmayr:2020:NTIRE2020Challenge}
A.~Lugmayr \emph{et~al.}, ``{{NTIRE}} 2020 {{challenge}} on {{real}}-{{world
  image super}}-{{resolution}}: {{methods}} and {{results}},'' in \emph{Proc.
  of the IEEE Conf. on Comput. Vis. and Patt. Recog. Work. (CVPRW)}, 2020, pp.
  494--495.

\bibitem{jacob:20:slr}
M.~Jacob, M.~P. Mani, and J.~C. Ye, ``Structured low-rank algorithms: theory,
  {MR} applications, and links to machine learning,'' \emph{{IEEE Sig. Proc.
  Mag.}}, vol.~37, no.~1, pp. {54--68}, Jan. 2020.

\end{thebibliography}
